\documentclass[a4paper,11pt]{article}

\usepackage[utf8]{inputenc}
\usepackage[T1]{fontenc}
\usepackage{graphicx,url}
\usepackage{a4wide}
\usepackage{mathpazo}
\usepackage[font=footnotesize,labelfont=bf]{caption} % Package used for the fix
\usepackage{amsmath}
\usepackage{bm}
\usepackage{amssymb}
\usepackage{multirow}
\usepackage{color,soul}
\usepackage{xurl}
\usepackage{authblk}

\usepackage[hidelinks]{hyperref}
\usepackage{listings}
\usepackage[numbers,super,sort&compress]{natbib}
\usepackage{nameref}

\usepackage{adjustbox}
\usepackage[shortlabels]{enumitem}
\setlist[enumerate]{nosep}

\usepackage[table,xcdraw]{xcolor}
\usepackage{tabulary}
\usepackage{booktabs}
\usepackage{longtable} 
\usepackage{array}     

\usepackage{titlesec}
\usepackage{lineno}
\definecolor{teal}{HTML}{1F9A8C}
\definecolor{softpink}{HTML}{F17CB0}
\definecolor{lightblue}{HTML}{6EC4E8}
\definecolor{purple}{HTML}{8C4DA1}
\definecolor{red}{HTML}{D62728}
\definecolor{blue}{HTML}{1F77B4}

\newcommand{\unit}[1]{\,\mathrm{#1}}

\titleformat{\paragraph}[runin]
  {\normalfont\bfseries}{\theparagraph}{1em}{} 
\titlespacing*{\paragraph}
  {0pt}{.5\baselineskip}{.5em}

\title{Cultural evolution of human beauty standards}

\author[1,2]{Louis Boucherie\thanks{Email: \href{mailto:louibo@dtu.dk}{louibo@dtu.dk}}}
\author[3,4]{Sagar Kumar}
\author[5,6]{Katharina Ledebur}
\author[2,7]{August Lohse}
\author[8]{Karolina \'{S}liwa}

\affil[1]{DTU Compute, Technical University of Denmark, Kongens Lyngby, Denmark}
\affil[2]{Center for Social Data Science, University of Copenhagen, Copenhagen, Denmark}
\affil[3]{Network Science Institute, Northeastern University, Boston, MA, USA}
\affil[4]{Center for Health Informatics Program, Boston Children's Hospital and Harvard Medical School, Boston, MA, USA}
\affil[5]{Complexity Science Hub, Vienna, Austria}
\affil[6]{Institute of the Science of Complex Systems, Medical University of Vienna, Vienna, Austria}
\affil[7]{Danish School of Education, University of Aarhus, Aarhus, Denmark}
\affil[8]{Vienna University of Economics and Business, Vienna, Austria}

\begin{document}

% FIX PART 1: Turn OFF list-of-figures recording for the Main Text
\captionsetup[figure]{list=no}
\captionsetup[table]{list=no}

% -----------------------------------------------------------------------------
% PART 1: MAIN TEXT
% -----------------------------------------------------------------------------

\maketitle

\begin{abstract}
Beauty standards shape self-perception and health through social comparison and objectification \cite{Festinger1954HBR,FredricksonRoberts1997POWQ}, while exposure to idealized imagery exacerbates body-image concerns \cite{grabe2008role,fioravanti2022exposure}.
Media and fashion are central arbiters of these ideals \cite{Entwistle2009AestheticEconomy}, yet long-term, quantitative, intersectional studies on how representation has changed remain scarce.
We assembled a dataset of 793{,}199 records spanning 25 years of advertising, magazine covers, runway shows, and editorials to quantify changes in anthropometric and demographic representation.
We find a paradox in the evolution of beauty ideals: while representational diversity has increased, the median model physique remains stable.
This is driven by selective plus-size inclusion at the upper tail, while the typical physique continues to diverge from the US population \cite{NHANES2021_2023}.
Intersectionally, non-white models are $4.5$ times more likely to be plus-size, indicating that progress in size inclusivity falls disproportionately on multiple underrepresented identities.
Stratifying the industry via a data-driven prestige hierarchy, we find that thinness is overrepresented at the top tier.
Finally, comparing two regulatory interventions \cite{Manifesto2006,FrenchLaw2017} we observe that numeric thresholds are more effective at reducing underweight appearances.
Our results quantify the cultural evolution in media and fashion, revealing that inclusion has increased;
however, gains are uneven and intersectionally concentrated on size and ethnicity, whereas the prevailing thin ideal remains largely unchanged.
\end{abstract}

\section*{Introduction}
The media and fashion industries promote narrow beauty ideals, emphasizing extremely thin female bodies and lean, muscular male physiques \cite{grabe2008role,mears2010size,jestratijevic2022body}.
Repeated exposure to these ideals in the contemporary hypermediated cultural landscape can have harmful consequences for self-perception through social comparison and objectification \cite{FredricksonRoberts1997POWQ,Festinger1954HBR}.
Viewers evaluate themselves against models who embody unrealistic ideals and typically feel they fall short \cite{want2009meta,Holland2016SNSBodyImage}.
As a result, body dissatisfaction increases, elevating the risk of psychological distress and disordered eating symptoms \cite{grabe2008role,barlett2008meta}, which can lead to elevated mortality \cite{Arcelus2011ArchGenPsych}.
These effects are consistent across genders and age groups \cite{Barnes2020PLoSONE,Sharpe2017ClinPsychSci}.
Despite these consequences, industry practices have normalized unsafe body sizes among professional models \cite{jestratijevic2022body}.
However, standards for beauty are not limited to body size alone.
Modeling hiring practices have shown to prioritize both thinness and racial exclusion \cite{mears2010size}.
More broadly, across media and cultural industries, skewed visibility normalizes the marginal status of underrepresented groups.
This legitimizes their exclusion from visibility, voice and opportunity in adjacent markets and institutions \cite{Tuchman1978Symbolic}.
Media representation also operates as a signifying practice: casting, styling and framing produce meaning by marking which bodies and identities are considered aspirational or professional.
This shapes hiring, taste formation and the allocation of attention and capital \cite{HALL1997Representation}.
Through cultivation, sustained exposure creates entrenched racial stereotypes \cite{dixon2019media}, and shifts shared beliefs about what kinds of people are typical or desirable \cite{MorganShanahan2010Cultivation}.
Together, these processes allow narrow representational regimes to reinforce social stratification while exporting psychological and health costs to the wider public.
Despite the public salience of these concerns, quantitative evidence capable of guiding policy and industry standards remains limited.
Prominent assessments rely on industry-commissioned audits \cite{VogueBusiness_SI_SS2025,VogueBusiness_SI_SS2024}, while academic studies are limited by small sample size \cite{mears2010size,hoppe2022microsociology}, region-specific focus \cite{jung2009cross,suradkar2024inclusivity,vranken2025no}, and an inability to track intersectional representation over time \cite{topaz2022race,bogar2024increased}.
As a result, we lack system-level estimates of how representation evolves across media channels and identities, and how far model physiques deviate from those of the general population.
%ours!
We address this gap by assembling a large, long-horizon dataset spanning the principal visual media sites where ideals have been historically established and disseminated -- advertisements, magazine covers, editorials and runway shows.
The dataset comprises 793{,}199 work records spanning 25 years (2000-2024), linking anthropometric measurements, visible traits (e.g., skin tone, hair/eye color), national-origin and images (see Methods Datasets) of models.
We leverage this dataset to quantify representation and body composition over time and across intersecting identities.
We provide system-level estimates that clarify where diversity has expanded, where it has stalled, and how far model physiques remain from general population.
Finally, to assess the effectiveness of regulations in bringing body representation closer to population norms, we compare the effects of two quasi-experimental policy interventions: the 2006 Milan BMI floor ($\geq 18.5\,\mathrm{kg\,m^{-2}}$) \cite{Manifesto2006} and the 2017 French medical certificate law \cite{FrenchLaw2017}.

\begin{figure}[h!]
    \centering
    \includegraphics[width=\textwidth]{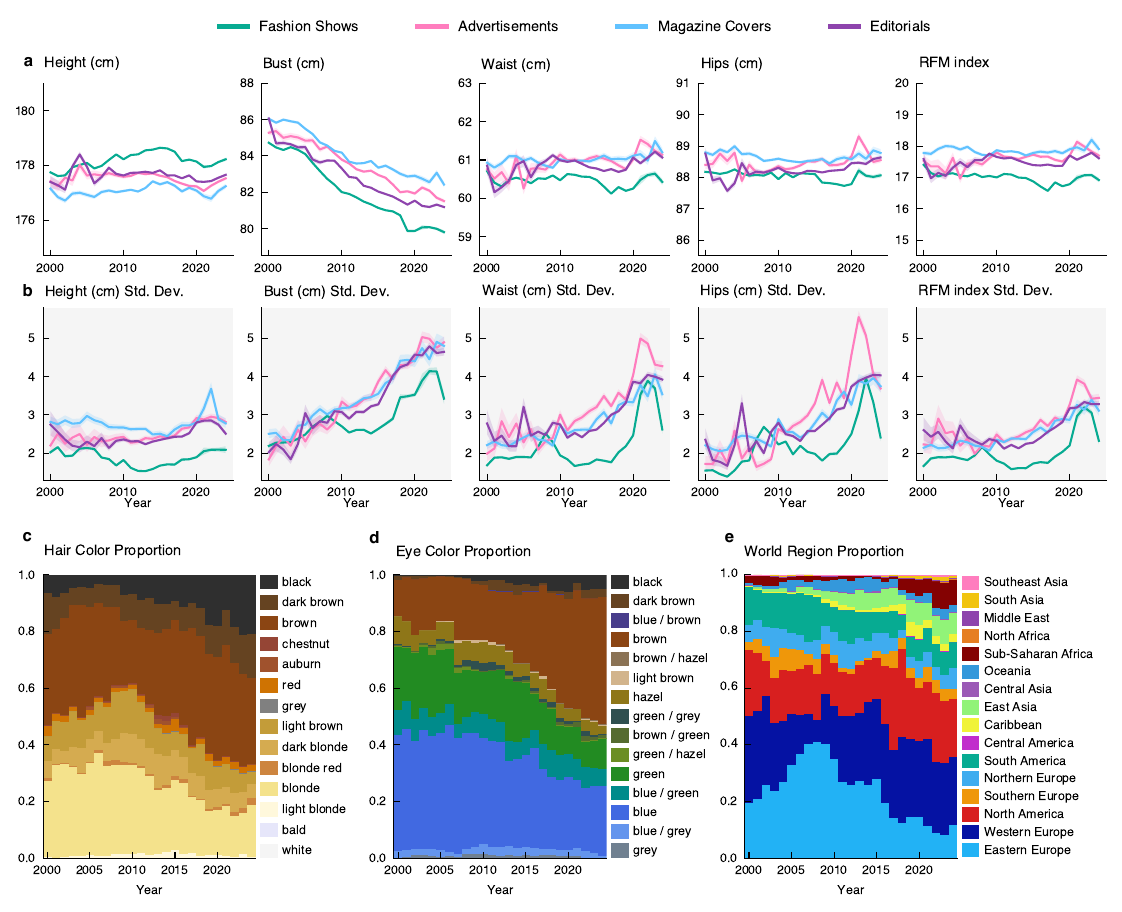}
    \caption{\textbf{Evolution of representational diversity for female models (2000--2024).}
    \textbf{(a)} Mean height, bust, waist, hips, and RFM (see Methods: RFM) by work type (\textcolor{teal}{fashion shows}, \textcolor{softpink}{advertisements}, \textcolor{lightblue}{magazine covers}, and \textcolor{purple}{editorials}).
    Anthropometric means remain stable across the period; only bust shows a modest decline. Shaded areas indicate 95\% confidence intervals.
    \textbf{(b)} Standard deviations of the same measures increase across all work types, indicating growing variability in represented body sizes despite stable central tendencies.
    \textbf{(c)} Hair color distribution shifts away from lighter phenotypes: blonde declines while darker shades gain share.
    \textbf{(d)} Eye color distribution shows similar diversification, with blue eyes declining and brown increasing.
    \textbf{(e)} National origins by world region: Eastern European representation peaks in the early 2000s then declines, while contributions from Sub-Saharan Africa, East Asia, and South Asia increase.
    See Fig.~\ref{fig:entropy} for entropy trends and Supplementary Sec.~\ref{si:worldregions} for region definitions.}
\label{fig:std_female_eu}
\end{figure}

\paragraph{Body measurement means remain stable while diversity in size, origin and appearance increases.}
Cross-sectional means for height, waist, hips, and Relative Fat Mass (Methods RFM) are constant across all work types, i.e. fashion shows, advertisements, magazine covers, and editorials for females (Fig.~\ref{fig:std_female_eu}a, Supplementary Results Male, Fig.~\ref{fig:si-evolution-male}a,b).
Only the bust circumference shows a consistent decline from an average of $85\unit{cm}$ to $82\unit{cm}$.
In contrast, dispersion increases significantly (Fig.~\ref{fig:std_female_eu}b, Supplementary Temporal trends). Since the early 2000s, the standard deviation of all measurements has roughly doubled (for example, hip circumference s.d. from $2\unit{cm}$ to $4\unit{cm}$), with similar patterns across work types, indicating a broader range of represented body sizes even as central tendencies remain stable.
As weight is not observed, we use RFM in place of body mass index (BMI);
RFM is a better proxy for body-fat level and avoids weight-height scaling biases \cite{woolcott2018relative,Woolcott2020}.

Appearances and origin traits diversify.
Hair and eye color distributions shift away from the dominance of lighter phenotypes: the prevalence of blonde hair and blue eyes declines while darker shades gain share (Fig.~\ref{fig:std_female_eu}c,d).
This coincides with a rise in diversity of national origins, as shown by increasing entropy over time (Fig.~\ref{fig:entropy}) with growing contributions from Sub-Saharan Africa, East Asia, and South Asia (Fig.~\ref{fig:std_female_eu}e).
Eastern European origins peak in the early 2000s and subsequently decline.
Western Europe and North America remain the largest sources, albeit with modest relative declines.
These trajectories are consistent with the gradual broadening of racialized features in fashion imagery reported in prior content analyses \cite{lindner2004images,baumann2008moral,reddy2018race}.

\begin{figure}[h!]
    \centering
    \includegraphics[width=\textwidth]{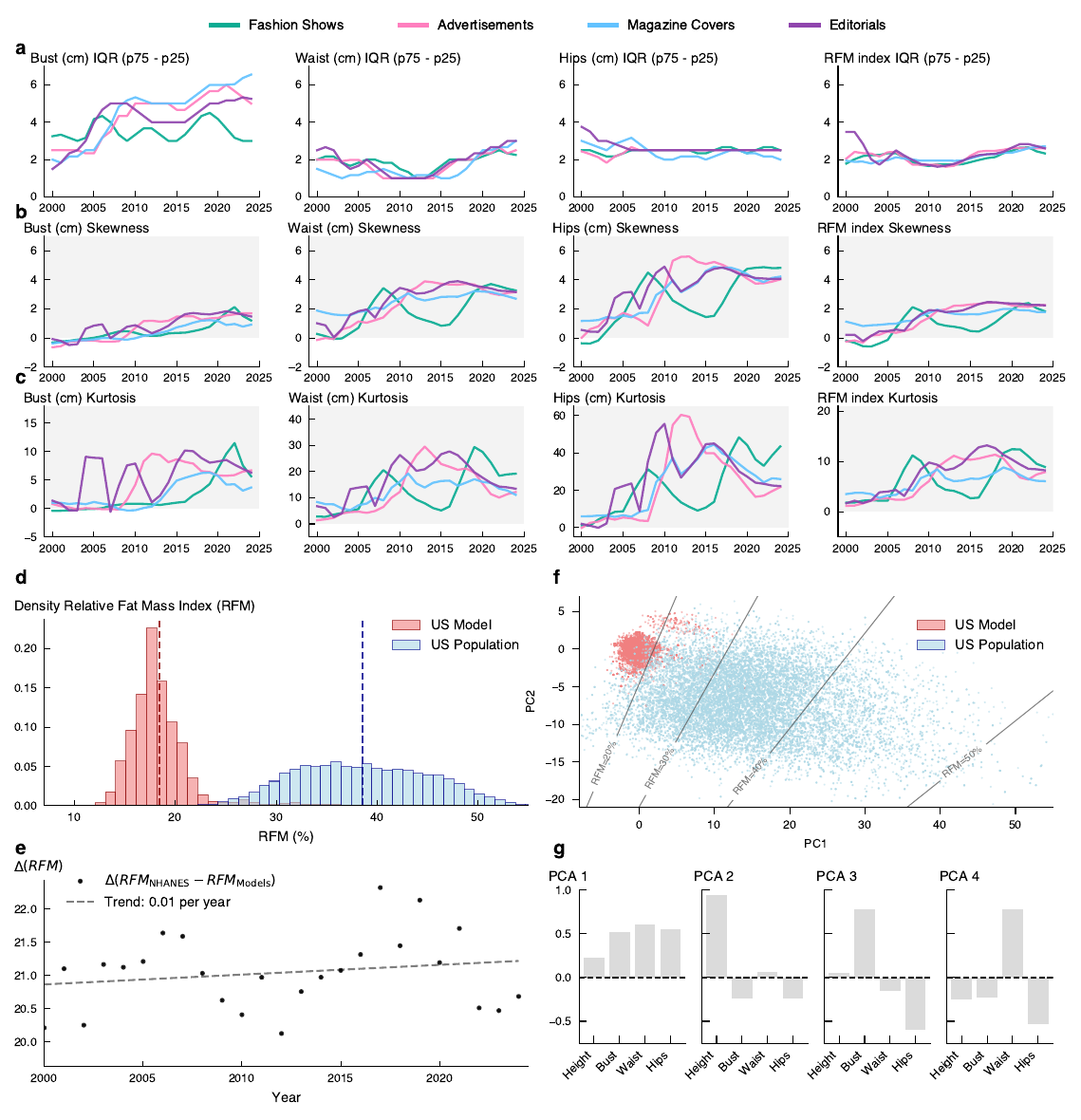} %pdf makes compialtion slow
    \caption{\textbf{Distribution shape and population benchmarking.}
    \textbf{(a)} Interquartile ranges (IQR) for bust, waist, hips, and RFM remain stable over time, indicating that the central 50\% of the distribution is unchanged.
    \textbf{(b)} Skewness shifts from near-zero to positive, indicating growing right-tail asymmetry (more large-size observations).
    \textbf{(c)} Excess kurtosis increases over time, reflecting heavier tails; estimates are noisy due to sensitivity to outliers.
    \textbf{(d)} RFM distributions for \textcolor{red}{US-based female models} and \textcolor{blue}{US women} aged 16--29 (NHANES \cite{NHANES2021_2023}) show minimal overlap;
    dashed lines indicate group means.
    \textbf{(e)} Annual gap $\Delta\mathrm{RFM} = \mathrm{RFM}_{\text{NHANES}} - \mathrm{RFM}_{\text{Models}}$ with fitted trend (${\sim}0.01$ percentage points per year), showing no convergence over two decades.
    \textbf{(f)} PCA projection separates models and population along RFM isolines, indicating selection primarily on body composition.
    \textbf{(g)} Component loadings: bust, waist and hips drive PC1; height dominates PC2.
    See Fig.~\ref{fig:pca} and Supplementary Multidimensional Analysis.}
    \label{fig:kurt_comparison_gen_pop}
\end{figure}

\paragraph{The increasing diversity is due to outliers, through the introduction of plus-size models.}
Distributional diagnostics show that the increase in variability is concentrated in the upper tail.
The interquartile ranges for bust, waist, hips, and RFM remain essentially flat (Fig.~\ref{fig:kurt_comparison_gen_pop}a, Supplementary Fig.~\ref{fig:si-forest-plot-iqr}), indicating stability in the median and bulk-physique.
By contrast, higher-moment statistics are increasing. Skewness is shifting from near zero to positive, and excess kurtosis is rising for size measures (Fig.~\ref{fig:kurt_comparison_gen_pop}b,c; Supplementary Fig.~\ref{fig:si-forest-plot-skew},\ref{fig:si-forest-plot-kurtosis}).
This indicates a heavier right tail with more large-size observations.
These patterns are consistent with a strategy of selective entry of plus-size models, which expands the tail while leaving central quantiles unchanged (Fig.~\ref{fig:body_tails}).
Hence, body-size “diversity” reflects the inclusion of a few outliers rather than a re-centering of the typical model body.

Male representation in media and fashion remains limited \cite{mears2010size,jestratijevic2022body}, a condition mirrored in our dataset, where male samples are smaller and more irregular than female ones (see Supplementary Results Male).
Despite this reduced coverage, the overall trend shows continuity rather than change.
Male models remain distinctly lean and muscular compared with the general population, and no meaningful broadening of represented body types is visible (Supplementary Fig.~\ref{fig:si-evolution-male},\ref{fig:si-iqr-male}).
This stability aligns with the recent media and industry reports noting the continued absence of size diversity in men's advertising and runway imagery \cite{VogueBusiness2026menswear}, and with psychological evidence that exposure to idealized physiques maintains narrow appearance norms \cite{barlett2008meta, Barnes2020PLoSONE,karsay2018sexualizing}.

\paragraph{Models remain physiologically distinct from the population.}
Comparisons with the general population show a clear separation in RFM (body-fat levels).
US-based female models cluster tightly around RFM of \(18.4 \pm 3.14\%\) with comparatively low variance.
In contrast, US women aged 16--29 in the NHANES study cluster near \(38.5 \pm 6.6 \%\) with a much broader spread (Fig.~\ref{fig:kurt_comparison_gen_pop}d; see Supplementary General Population).
The two distributions exhibit minimal overlap, indicating that models remain concentrated at the low-RFM extreme (underweight range).
The annual gap, \(\Delta\mathrm{RFM} \equiv \mathrm{RFM}_{\text{NHANES}} - \mathrm{RFM}_{\text{Models}}\) is large (about 20 percentage points) and temporally stable, with only a negligible linear increase of \({\sim}0.01\) percentage points per year (Fig.~\ref{fig:kurt_comparison_gen_pop}e).
Thus, the physiological distance between the models and age-matched women has not meaningfully narrowed or widened over the past two decades.
To account for multivariate body shape, we perform principal components analysis (PCA) using height, bust, waist and hips (Fig.~\ref{fig:kurt_comparison_gen_pop}f,g).
In the PC1--PC2 plane, groups separate along a direction orthogonal to RFM isolines, indicating selection primarily on RFM (body-fat level).
The decomposition shows that PC1 captures overall size (positive on height and circumferences), whereas PC2 indexes slenderness (positive on height, negative on circumferences) (Fig. \ref{fig:pca}, Supplementary Multidimensional Analysis).
Together, these results show a strong selection; models exhibit an exceptionally lean, low-variance phenotype that differs significantly from that of age-matched women.
This disparity persists despite the diversity gains documented in Fig.~\ref{fig:std_female_eu}.
Having established that diversity gains are concentrated in distributional tails rather than central tendencies, we next examine whether these patterns are uniform across the industry or stratified by institutional position.

\begin{figure}[h!]
    \centering
    \includegraphics[width=0.6\textwidth]{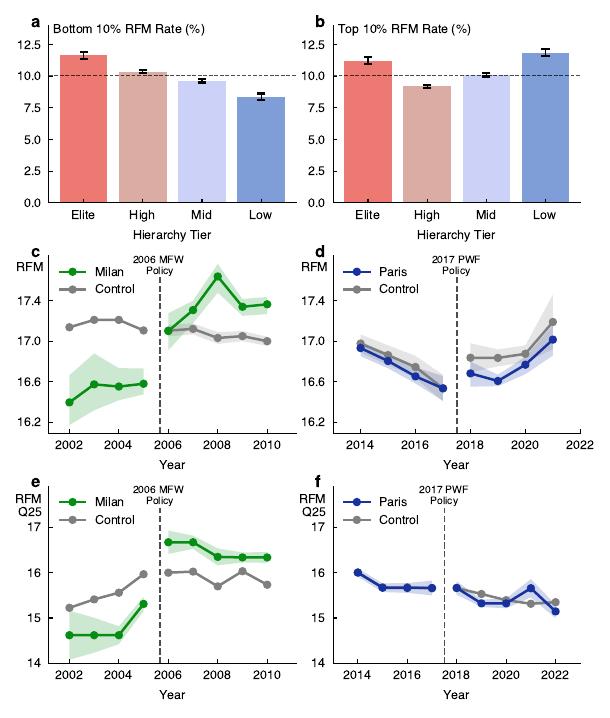}
    \caption{\textbf{Prestige-tier dependence and policy impacts on model Relative Fat Mass (RFM).}
    \textbf{(a--b)} Industry stratification by brand prestige.
    Brands are assigned to four tiers (Elite, High, Mid, Low) on the model--brand bipartite network (see Methods: Network Hierarchy).
    \textbf{(a)}~The share of low-RFM models (bottom decile of the industry-wide distribution) declines monotonically from Elite to Low tiers, indicating that thinness selection intensifies at the top of the hierarchy.
    \textbf{(b)}~The share of high-RFM models (top decile) follows a U-shaped pattern: over-represented at both Elite and Low tiers, with the smallest shares in the middle tiers---suggesting that mid-tier brands are less willing to cast plus-size models than either elite houses or lower-prestige outlets.
    \textbf{(c--f)} Policy impact analysis using difference-in-differences. 
    \textbf{(c)}~Mean RFM at Milan Fashion Week (green) versus control cities (grey), 2002--2010.
    Following the 2006 BMI floor ($\geq$18.5\,kg/m$^2$), Milan exhibits a discrete upward shift of approximately 0.5 RFM units relative to the stable control trend.
    \textbf{(d)}~Mean RFM at Paris Fashion Week (blue) versus control cities, 2013--2022. The 2017 French medical-certificate law produces no visible discontinuity;
    both Paris and control cities experience parallel increases coinciding with rising plus-size representation industry-wide (cf.\ Fig.~\ref{fig:intersectional}).
    \textbf{(e)}~The 25th percentile of RFM at Milan jumps by more than one unit after 2006, confirming that the policy effect is concentrated in the lower tail of the distribution.
    \textbf{(f)}~The 25th percentile of RFM at Paris shows no differential change relative to control cities following the 2017 law.
    Shaded bands indicate 95\% confidence intervals. Vertical dashed lines mark policy implementation.}
    \label{fig:tier_policy}
\end{figure}

\paragraph{Effects are heterogeneous across the industry.}
To test whether diversity evolves uniformly across fashion and media, we stratify brands and magazines by prestige.
We build a bipartite collaboration network linking models to brands or magazines and compute centrality-based influence scores to derive a data-driven hierarchy \cite{Clauset2015SciAdv,Kleinberg1999}.
The distribution is partitioned into ordered tiers (Elite, High, Mid, Low; see Methods Network Hierarchy \nameref{sec:network}).
This hierarchy captures sustained influence and aligns with expert assessments of brand and magazine status (Fig.~\ref{fig:network}; Supplementary Network Hierarchy).
Using these tiers, we compare the prevalence of models at the extremes of RFM.
The share of low-RFM appearances (bottom decile) declines monotonically from Elite to Low, indicating stronger thinness selection at the top of the hierarchy (Fig.~\ref{fig:tier_policy}a).
By contrast, high-RFM appearances (top decile) follow a U-shaped pattern, over-represented at both Elite and Low tiers, and least common in the middle tiers (Fig.~\ref{fig:tier_policy}b).
This suggests that, while brands and magazines in the middle tiers are not as thin-selective as those in the Elite tier, they are also less willing to hire plus-size models.

\begin{figure}[htb]
    \centering
    \includegraphics[width=0.6\linewidth]{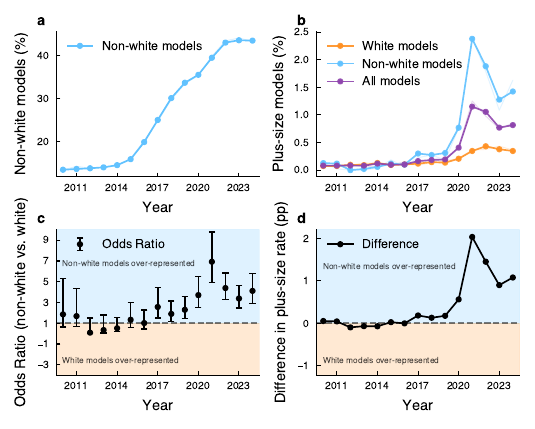}
    \caption{\textbf{Intersectional minorities carry visible diversity}
    \textbf{(a)} Evolution of the proportion of non-White models (Methods: Ethnicity Attribution).
    \textbf{(b)} Evolution of the proportion of plus-size model by ethnicity (White and non-White) 
    (Methods: Plus-size Attribution). 
    \textbf{(c)} Odds ratio of being a plus-size model for the two populations, It represent how more likely are non-White models to be plus-size compare to White models.
    \textbf{(d)} Difference between the proportion of plus-size model per ethnicity.
    Non-White models are almost always over-represented among the plus sized models (Methods Odds ratio and intersectionality analysis)}
    \label{fig:intersectional}
\end{figure}

\begin{figure}[htb]
    \centering
    \includegraphics[width=0.9\linewidth]{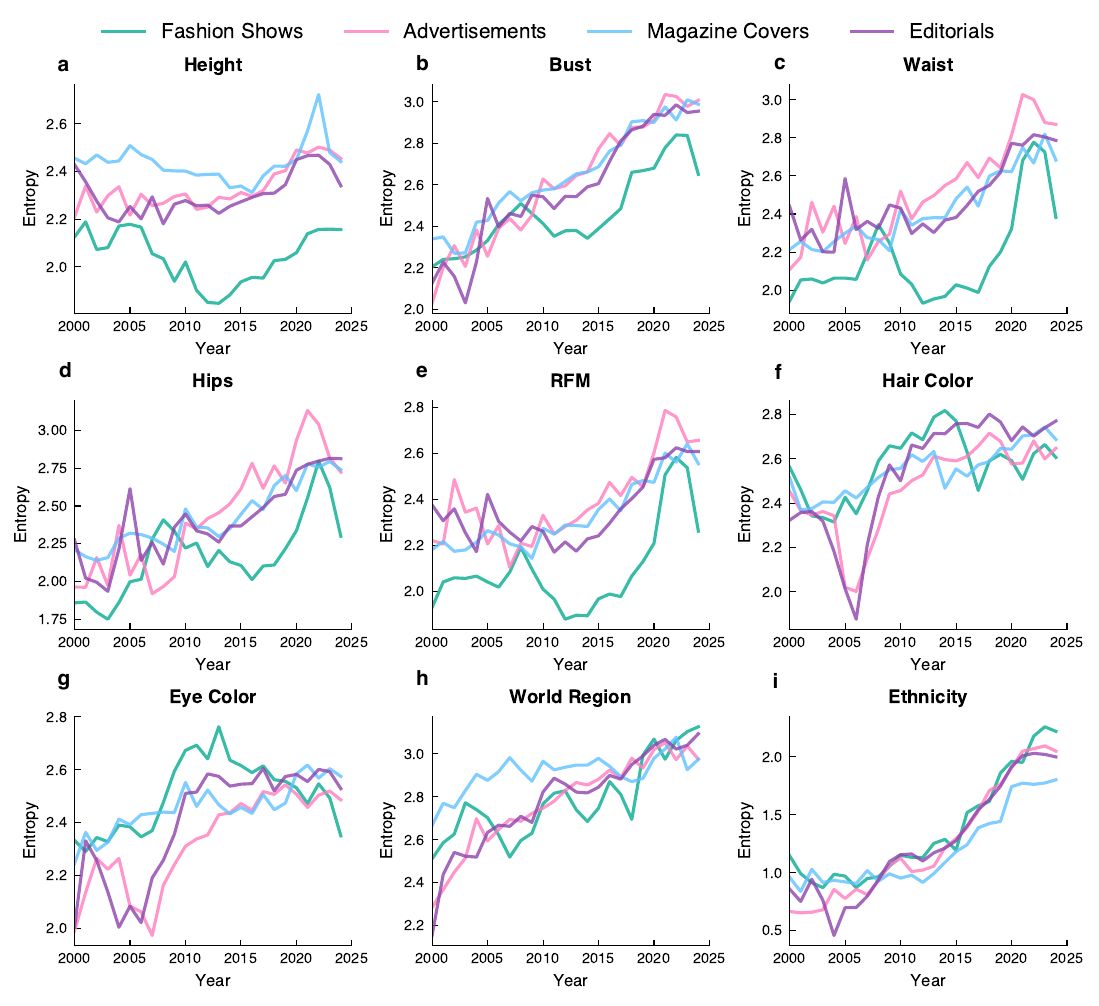}
    \caption{\textbf{Temporal evolution of diversity across body measurements and demographic characteristics}
      Entropy measures quantify diversity for nine variables across four career event types: \textcolor[HTML]{07AB92}{fashion shows}, \textcolor[HTML]{FF7DBE}{advertisements}, \textcolor[HTML]{61C2FF}{magazine covers}, and \textcolor[HTML]{8E44AD}{editorials} (2000-2024).
      \textbf{(a-c)} Body size measurements (height, bust, waist) show stable differential entropy with modest temporal variation.
      \textbf{(d-e)} Hip and RFM entropy trends reveal consistent diversity patterns across all fashion contexts.
      \textbf{(f)} Hair color Shannon entropy demonstrates stable categorical diversity over time.
      \textbf{(g)} Eye color diversity shows slight increasing trends in recent years for some event types.
      \textbf{(h)} Geographic diversity (world region) exhibits distinct patterns between event types, with fashion shows maintaining higher entropy than other contexts.
      \textbf{(i)} Ethnicity diversity shows temporal increases across all event types.}
    
\label{fig:entropy}
\end{figure}

\begin{figure}[htb]
    \centering
    \includegraphics[width=1\linewidth]{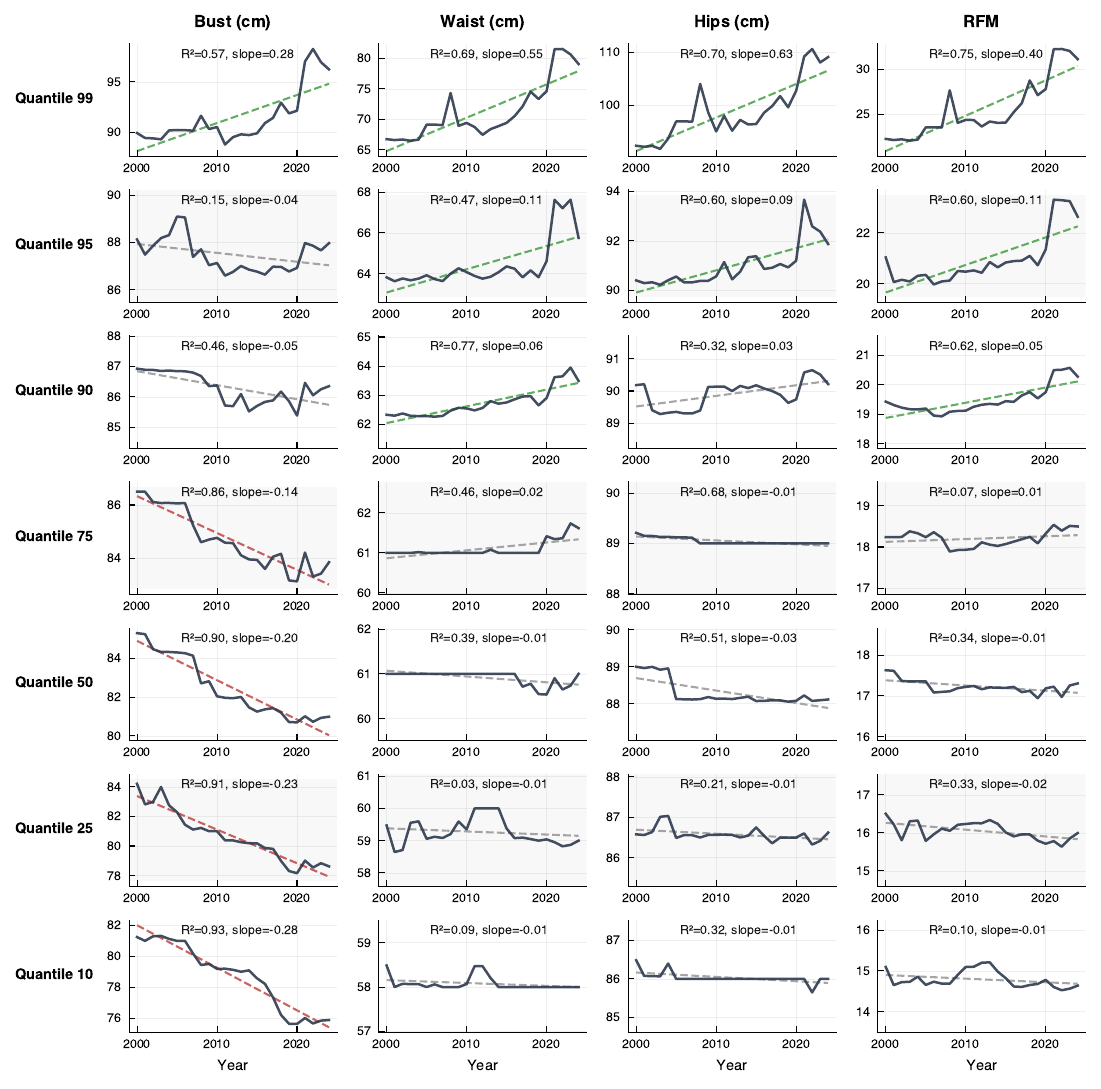}
    \caption{\textbf{Temporal Evolution of Body Measurement Quantiles (2000--2024).}
  This figure displays the temporal trends of seven quantiles (99th, 95th, 90th, 75th, 50th, 25th, and 10th percentiles, from top to bottom) across four body measurements (bust, waist, hips in cm, and Relative Fat Mass index) for
  female fashion models.
  Each panel shows the values calculated across fashion shows, advertisements, magazine covers, and editorials.
  Solid dark lines represent the year-by-year quantiles.
  Dashed lines indicate linear regression fits, colored green for positive slopes (increasing trends) and red for negative slopes (decreasing trends).
  }
\label{fig:body_tails}
\end{figure}

\begin{figure}[htb]
    \centering
    \includegraphics[width=1\linewidth]{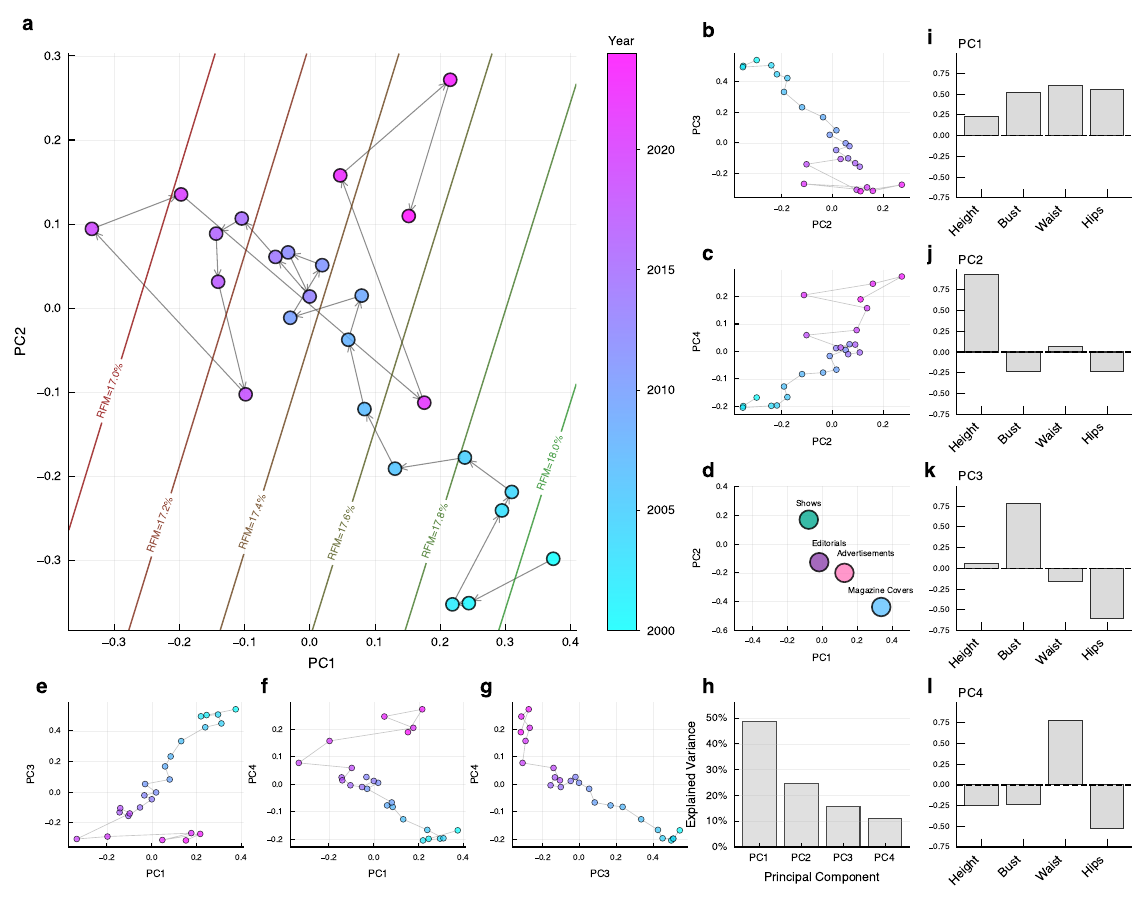}
      \caption{\textbf{Temporal evolution of body measurements in PCA space}
      \textbf{(a)} Main trajectory: PC1-PC2 showing yearly means from 2000-2024 with directional arrows and RFM isolines (17.0-18.0\%).
      Colormap indicates temporal progression from early (blue) to recent years (pink).
      \textbf{(b-c)} Secondary projections: PC2-PC3 and PC2-PC4 reveal additional temporal patterns beyond the primary PC1-PC2 plane.
      \textbf{(d)} Career event averages: distinct positioning of \textcolor{teal}{fashion shows}, \textcolor{softpink}{advertisements}, \textcolor{lightblue}{magazine covers}, and \textcolor{purple}{editorials} in PC1-PC2 space demonstrates body measurement variation across fashion and media contexts.
      \textbf{(e-g)} Complementary PC combinations (PC1-PC3, PC1-PC4, PC3-PC4) showing consistent temporal trajectories across all principal component dimensions.
      \textbf{(h)} Explained variance: PC1-PC4 account for cumulative variance with PC1 and PC2 dominating.
      \textbf{(i-l)} Component loadings: waist and hips drive PC1 (body composition), height dominates PC2 (stature), bust and waist contribute to PC3-PC4 (secondary shape variation).
      (see Supplementary Multidimensional Analysis).}
\label{fig:pca}
\end{figure}

\begin{figure}[htb]
    \centering
    \includegraphics[width=1\linewidth]{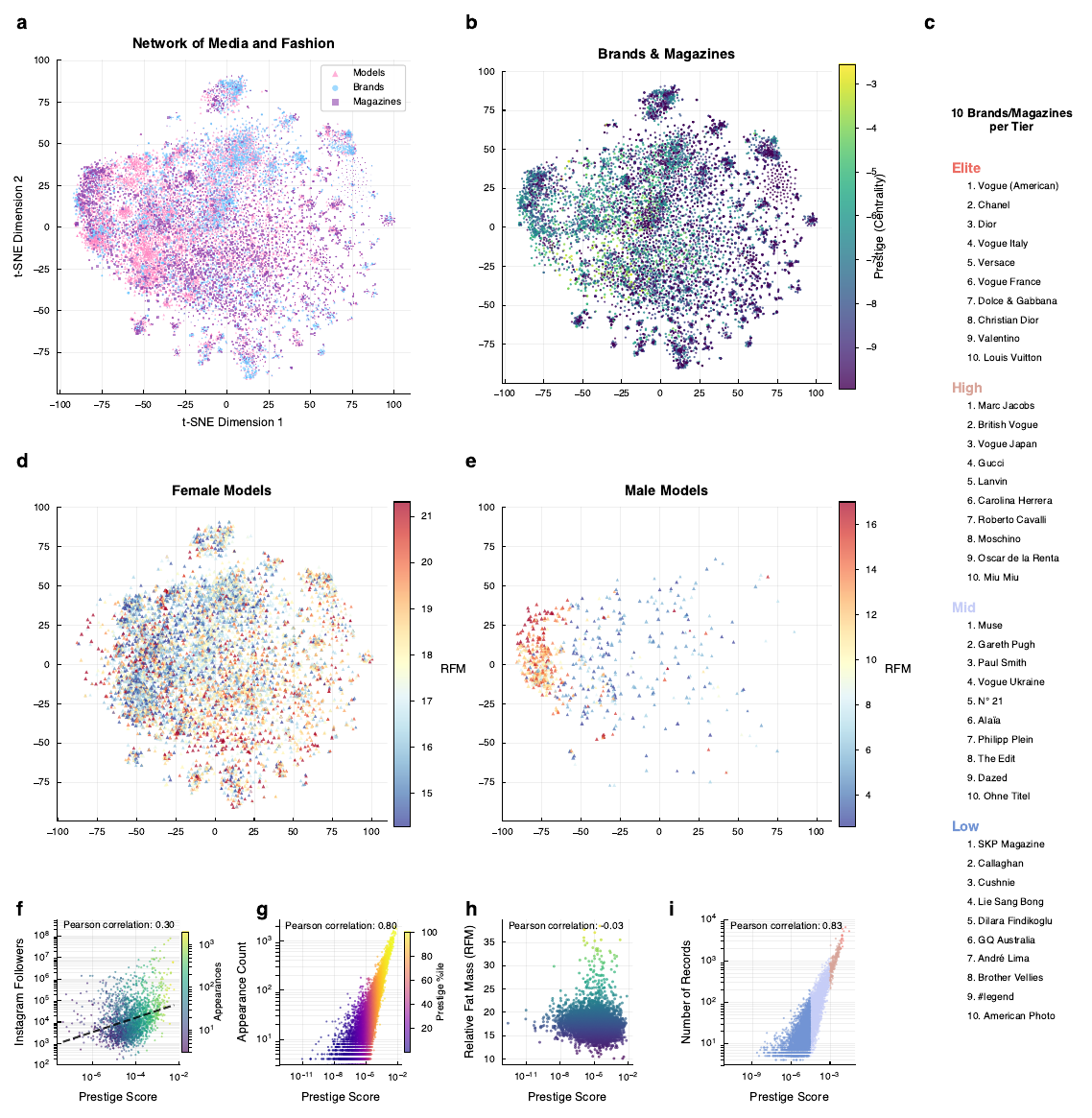}
    \caption{\textbf{Network structure and prestige stratification in the fashion industry.}
\textbf{(a)} Bipartite network embedding of models, brands, and magazines using node2vec then t-SNE.
\textbf{(b)} Brands and magazines colored by prestige (centrality score).
\textbf{(c)} Top 10 brands/magazines per prestige tier (Elite, High, Mid, Low).
\textbf{(d)} Female models colored by RFM (see Methods; RFM).
\textbf{(e)} Male models colored by RFM (see Methods; RFM).
\textbf{(f)} Model prestige vs.\ Instagram followers (log–log scale; Pearson $r=0.89$).
\textbf{(g)} Model prestige vs.\ appearance count (log–log scale; Pearson $r=0.83$).
\textbf{(h)} Female model prestige vs.\ RFM (Pearson $r=-0.03$), colors as in \textbf{(g)}.
\textbf{(i)} Brand prestige vs.\ number of show records (log–log scale; Pearson $r=0.30$).
}
\label{fig:network}
\end{figure}

\begin{figure}[htb]
    \centering
    \includegraphics[width=1\linewidth]{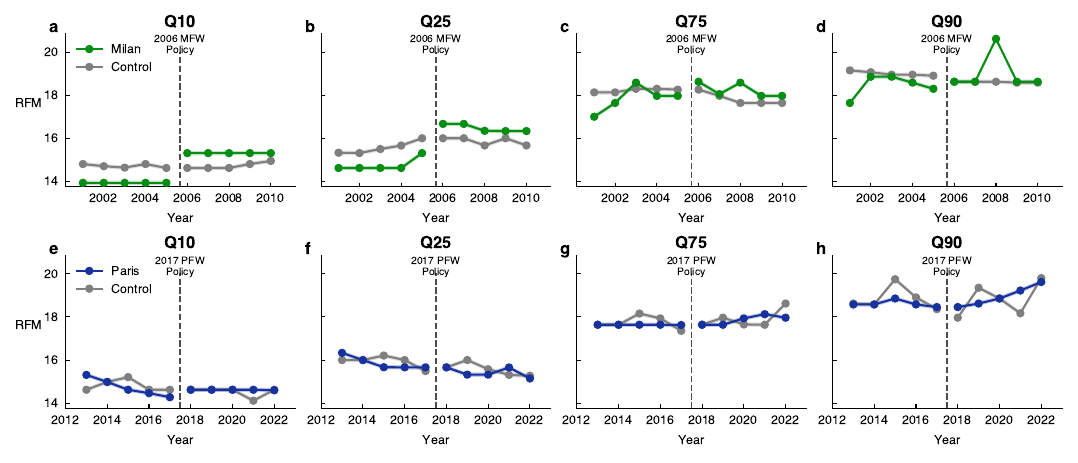}
    \caption{\textbf{Quantile effects of policy shocks on model RFM (body-fat levels).}
\textbf{(a-d) Milan (2002--2010).} Quantile trends at $Q10$, $Q25$, $Q75$ and $Q90$ compare Milan Fashion Week (MFW) with the rest of the industry, with the 2006 policy marked.
The discontinuity is concentrated in the lower quantiles ($Q10,Q25$), which jump upward at the policy date, while upper quantiles ($Q75,Q90$) show little change-consistent with lower-tail pruning.
\textbf{(e-h) Paris (2012-2022).} Quantile trends around the 2017 Paris Fashion Week (PFW) medical-certificate rule show that lower quantiles continue to decline after the policy, whereas upper quantiles rise.
Both Paris and the rest of the industry shift upward, with the industry-wide increase exceeding that of Paris, indicating that mean gains in RFM are driven by growth in the number of high-RFM models (greater plus-size representation) rather than a decrease of low-RFM models.
}
\label{fig:quantile_policy}
\end{figure}

\begin{figure}[htb]
    \centering
    \includegraphics[width=1\linewidth]{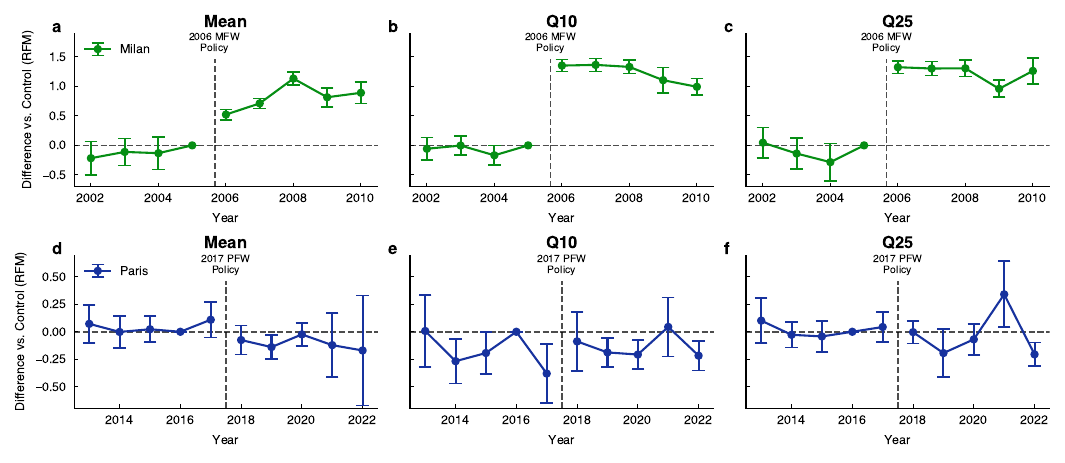}
    \caption{\textbf{Event Study}
\textbf{(a-c) Milan (2002--2010).} Difference in RFM between control and Milan (treated city) for mean, $Q10$ and $Q25$.
We compare Milan Fashion Week (MFW) with the rest of the industry, with the 2006 policy marked.
The discontinuity is concentrated in the lower quantiles ($Q10,Q25$), which jump upward at the policy date.
\textbf{(d-f) Paris (2012-2022).} Difference in RFM between control and Paris (treated city) for mean, $Q10$ and $Q25$.
After the 2017 medical-certificate rule, there is no significant difference between in mean or lower quantiles, indicating no clear policy-driven shift relative to the industry.
}
\label{fig:event_study}
\end{figure}

\paragraph{Policy shocks reveal how regulation affects body-size representation.}
To evaluate the impact of regulatory interventions on body-size representation, we study two quasi-experimental policy shocks.
In 2006, the Milan fashion week introduced a ban on female models with a body mass index (BMI) below 18.5 \cite{Manifesto2006}.
In 2017, France introduced a medical-certificate requirement for models, without specifying numeric cutoffs \cite{FrenchLaw2017}.
We estimate the effect of each intervention using a two-way fixed-effects (TWFE) difference-in-differences event-study framework (see Methods: Difference-in-differences policy analysis).
For Milan, we compare RFM trends for female in the treated city to a control group comprising New York, Paris, London, and other major fashion capitals over 2002--2010, with 2005 as the reference year.
For Paris, we use an analogous design over 2013--2022, with 2016 as the reference year.
Milan's BMI floor produced measurable effects. Event-study coefficients reveal a discrete, sustained increase in model body size following the 2006 intervention (Fig.~\ref{fig:tier_policy}c).
The 2006 post-policy coefficient indicates that the regulation increased mean RFM by $0.52\pm0.045$ unit relative to the counterfactual trend (Fig.~\ref{fig:event_study}a).
Crucially, this shift was concentrated in the lower tail: quantile DiD estimates show that the 10th and 25th percentiles of RFM rose by $1.35\pm0.052$ and $1.33\pm 0.055$ units, respectively ((Fig.~\ref{fig:tier_policy}e), Table~\ref{tab:milan_main_results}; Fig.~\ref{fig:event_study}b-c).
This pattern indicates that the numeric threshold successfully reduced the prevalence of underweight models rather than shifting the entire distribution.
Pre-policy event-study coefficients are small in magnitude and jointly indistinguishable from zero (Wald test: $\chi^2 = 3.53$, $p = 0.317$ for the mean; see Table~\ref{tab:milan_main_results}), supporting the parallel-trends assumption.
Placebo tests assigning treatment to non-intervention years (2004, 2008) yield null results (Supplementary Policy Analysis).
France's medical-certificate law showed no detectable effect. By contrast, the 2017 French regulation, which required health certificates but imposed no explicit BMI threshold, produced no systematic deviation from control trends in the event-study profiles (Fig.~\ref{fig:tier_policy}d; Table~\ref{tab:paris_main_results}).
Quantile diagnostics show stable lower tails, indicating that any mean increases in RFM during this period reflect the industry-wide rise in plus-size representation ((Fig.~\ref{fig:tier_policy}f, Fig.~\ref{fig:quantile_policy}e--h; Fig.~\ref{fig:intersectional}) rather than policy-induced reductions in underweight appearances. Taken together, these results suggest that explicit numeric thresholds are more effective than flexible guidelines at reducing underweight representation \cite{kaplow2013rules}.

\paragraph{Intersectional minorities carry body diversity.} 
We showed that body-size diversity has increased through the selective inclusion of plus-size outliers.
In the meantime, diversity in phenotypes and national origins has also expanded.
We now examine whether these dimensions of inclusion are distributed independently or intersect on the same bodies.
Representation of non-White and plus-size models have increased substantially since the early 2010s.
The share of non-White models rose from 13\% in 2011 to 44\% in 2024 (Fig.~\ref{fig:intersectional}a).
Plus-size models, initially almost absent, expanded to 1\% over the same period (Fig.~\ref{fig:intersectional}b).
However, gains in size diversity are disproportionately carried by non-White models.
Starting from similar levels in 2011, the gap in representation between White and non-White models has since widened (Fig.~\ref{fig:intersectional}b).
Since 2017, the odds ratio of being a plus-size model is significantly higher for non-White models.
It averages 4.5 for the 2017-2024 period, meaning that non-White models are 4.5 times more likely to be plus-size models (Fig.~\ref{fig:intersectional}c-d, Table~\ref{tab:intersectional}, Methods: Odds ratio).
This intersectional concentration shows that much of the industry’s diversity is disproportionately borne by non-White models reflecting showcase inclusion rather than systemic change \cite{topaz2022race, reddy2018race,foster2022model}.
This imbalance is not new. Slenderness and ethnicity have long been decisive criteria in model selection, both within fashion \cite{mears2010size} and in the industry as a whole \cite{doyle2023diversity}.
Research on performative diversity argues that industries often favor intersectional representation that maximize symbolic value while minimizing structural change \cite{lee2020all,crenshawmapping1991}.
In this light, body-size inclusion is disproportionately borne by racialized models, allowing brands to satisfy multiple representational pressures simultaneously by consolidating multiple markers of diversity within a single body.

\section*{Discussion}

Our findings reveal a paradox in the cultural evolution of beauty standards.
Although media and fashion representation has diversified in terms of visible traits and national origins, physique ideals remain stable.
Variation in body size reflects the selective inclusion of outliers rather than a redefinition of the standard model body type.
This symbolic inclusivity may accentuate, rather than diminish, the contrast between dominant thinness and marginal diversity \cite{cavusoglu2023extending,gruys2022fit}.
Diversity is also unevenly distributed. Increases in body-size representation are disproportionately carried by non-White individuals, indicating that the labor of inclusivity falls on groups already marginalized by ethnicity and region.
This intersectional concentration provides quantitative evidence for critiques of superficial inclusivity and aesthetic labor in fashion \cite{topaz2022race,reddy2018race,foster2022model}.
When benchmarked against the general population, this selective diversification reveals its limits: for two decades, the gap in body composition between US models and US women has remained close to 20 percentage points, with no sign of convergence.
Symbolic diversification has not narrowed the divide between fashion’s ideals and real bodies, a disconnect long associated with body dissatisfaction and disordered eating \cite{grabe2008role,want2009meta}.
These results have practical implications. The Milan BMI rule produced a measurable rise in model body-fat levels, whereas the French medical-certificate law had no clear effect, suggesting that explicit numerical thresholds work better than flexible guidelines \cite{kaplow2013rules}.
Yet such measures address only the lowest end of representation.
Rebalancing ideals requires attention to the center as well as the tails, encouraging broader casting, transparent reporting of body-measurement statistics, and independent monitoring of diversity commitments.
Policies that reward gradual shifts in typical physiques, rather than occasional extremes, could move the industry toward more realistic and inclusive norms.
Several limitations should be acknowledged. 
For 23\% of records where self-reported profile fields were unavailable, gender and ethnicity were imputed using probabilistic classifiers (face-based), which provide scalable but imperfect proxies for identity, the remaining 77\% use labels self-reported by models or agencies.
These methods necessarily impose simplified categories. Gender is reduced to a binary frame, and ethnicity is inferred from appearance rather than self-identification.
Such simplifications risk excluding or misclassifying transgender, non-binary, mixed-ethnicity, and other marginalized groups, and they collapse distinct identities into broad “White/non-White” categories.
Although these tools enable systematic analysis across a large dataset, they should not be read as capturing lived identity in its complexity.
While this dataset offers one of the most comprehensive records of model representation to date spanning anthropometric, phenotypic, and demographic attributes, it remains incomplete.
Weight and age are disclosed only for a subset of models due to the industry's historical reluctance to disclose health-related metrics \cite{jestratijevic2022body}.
Lacking direct weight data, we use Relative Fat Mass (RFM) as a proxy derived from height, waist, and sex.
Although RFM is generally more accurate than BMI for estimating body fat \cite{woolcott2018relative}, it is not equivalent to a clinical measurement.
The narrow beauty standards documented here are precisely those encoded in generative AI training sets.
As virtual and AI-generated models enter advertising and editorial contexts \cite{rufo2025vogueai}, these systems risk amplifying rather than correcting representational bias \cite{vargas2025visual,castleman2025adultification}.
Mitigating algorithmic harm will require systematic auditing of training corpora, fairness benchmarks for generated outputs, and transparency standards governing synthetic media in commercial contexts \cite{kalluri2025computer}.

\section*{Methods}
\small{
\subsection*{Datasets}
\label{sec:Datasets}

We compiled a dataset from professional pages on \url{https://models.com} and \url{https://www.fashionmodeldirectory.com}.
These platforms aggregate portfolio data submitted by modeling agencies and verified by industry professionals;
profile attributes are typically self-reported by models or provided by their representing agencies.
After de-duplication at the name-agency-URL level, the corpus comprises 793{,}199 records.
Available attributes include anthropometrics (height, bust, waist, hips, dress size, shoe size), identity fields (name, nationality; when available, country of origin, age, weight, gender, ethnicity), Instagram handle, and profile photographs.
All anthropometric measurements were standardized to SI units. The records are subdivided in the following categories: Advertisements (92{,}283 records), Magazine Covers (86{,}382), Fashion Runways (413{,}265) and Editorials (146{,}503).
Each record includes the model identifier, job type, and at least one associated image.
These images are used for downstream annotations. When self-reported profile fields are unavailable (23\% of the record), gender and ethnicity are estimated using probabilistic classifiers applied to profile photographs, these serve as scalable but imperfect proxies for identity (see Supplementary Gender and Ethnicity Attribution) \cite{karkkainen2021fairface}.
Instagram follower counts were downloaded from \url{https://www.instagram.com/} using model handles listed on the professional pages.
To compare body measurements of American models to a general population sample, we use data on demographics and body measurements from the National Health and Nutrition Examination Survey (NHANES) cycle~\cite{NHANES2021_2023} (2000-2023) (\href{https://www.cdc.gov/nchs/nhanes/}{https://www.cdc.gov/nchs/nhanes/}).
We select only individuals between 16 and 29 to approximate the age range of the models.
We use the information on gender, height and waist measurements to calculate the Relative Fat Mass (RFM) metric.

\subsection*{Relative Fat Mass (RFM)}
\label{sec:RFM}

Relative Fat Mass (RFM) is a body composition index that estimates an individual's body fat percentage using height, waist circumference, and gender which was found to better predict whole-body fat percentage among adult individuals than the body mass index (BMI)~\cite{woolcott2018relative}.
It has been found to be a more accurate predictor of mortality and to reduce the risk of underdiagnosing obesity~\cite{Woolcott2020}.
It is defined as: 
\begin{equation}
\label{eq:rfm}
\text{RFM} = 64 - 20 \times \left( \frac{\text{height}}{\text{waist}} \right) + 12 \times \text{sex}, \quad \text{where sex} =
\begin{cases}
0 & \text{for men} \\
1 & \text{for women}
\end{cases}
\end{equation}

\subsection*{Statistical methods}
All error bars throughout this paper correspond to bootstrapped 95\% confidence intervals.
When these bootstrapped 95\% confidence intervals do not include the null value (typically 0), they indicate a statistically significant difference at the $\alpha=0.05$  level.

\subsection*{Network analysis of media and fashion prestige hierarchy}
\label{sec:network}

We model the brand--model relation as a weighted bipartite graph
\(G=(\mathcal{M},\mathcal{B},\mathcal{E})\), where \(\mathcal{M}\) and \(\mathcal{B}\) denote models and brands (fashion houses and magazines), respectively.
An edge \(e_{mb}\in\mathcal{E}\) carries weight \(w_{mb}\in\mathbb{R}_{+}\), the number of appearances of model \(m\) for brand \(b\) during the study period.
The weighted degree of node \(v\) is
\[
k_v=\sum_{u\in\mathcal{N}(v)} w_{vu},
\]
and correspond to the number of works for a model or the number of event for a brand.
To quantify hierarchical influence, we compute eigenvector centrality on one-mode projections with fractional co-appearance weights (each event contributes unit mass across co-participants) \cite{Clauset2015SciAdv}.
For the model projection \(G_{\mathcal{M}}\),
\[
\tilde w_{mm'}=\sum_{b\in\mathcal{B}}\frac{w_{mb}\,w_{m'b}}{\sum_{m''\in\mathcal{M}} w_{m''b}}\!,
\]
and centralities \(c_v\) satisfy
\[
c_v=\lambda^{-1}\sum_{u\in\mathcal{N}_{\mathcal{M}}(v)} \tilde w_{vu}\,c_u,
\qquad \text{normalized so } \sum_{v\in\mathcal{M}} c_v=1.
\]
Brand prestige is computed analogously on the brand projection \(G_{\mathcal{B}}\) with
\[
\tilde w_{bb'}=\sum_{m\in\mathcal{M}}\frac{w_{mb}\,w_{mb'}}{\sum_{b''\in\mathcal{B}} w_{mb''}}\!.
\]

Tiering by appearance-weighted percentiles.
Because elite actors generate a disproportionate share of appearances, tier cutpoints are defined on appearance-weighted (edge-weighted) centrality distributions rather than on unweighted actor counts.
Let \(s_v:=k_v\) denote the appearance volume of actor \(v\). Define the weighted CDF
\[
F^{(s)}_c(t)=\frac{\sum_{v:\,c_v\le t} s_v}{\sum_{v} s_v},
\]
and the weighted quantile \(Q^{(s)}_{p}(c)=\inf\{t:\,F^{(s)}_c(t)\ge p\}\).
Actors are assigned to four ordered tiers by weighted percentile bands (share of total appearances):
\[
\mathcal{T}_{\mathrm{Elite}}=\{v:\,c_v\ge Q^{(s)}_{0.90}(c)\},\quad
\mathcal{T}_{\mathrm{High}}=\{v:\,Q^{(s)}_{0.50}(c)\le c_v<Q^{(s)}_{0.90}(c)\},
\]
\[
\mathcal{T}_{\mathrm{Mid}}=\{v:\,Q^{(s)}_{0.10}(c)\le c_v<Q^{(s)}_{0.50}(c)\},\quad
\mathcal{T}_{\mathrm{Low}}=\{v:\,c_v<Q^{(s)}_{0.10}(c)\},
\]
corresponding to 90-100 (Elite), 50-90 (High), 10-50 (Mid), and 0-10 (Low) percentiles by appearance mass.

\subsection*{Difference-in-differences policy analysis}

We estimate the impact of the 2006 BMI regulation in Milan using a two-way fixed-effects difference-in-differences (DiD) design with Milan as the treated city and other major fashion capitals (New York, Paris, London, and additional fashion weeks) as controls over the period 2002--2010.
We apply an analogous specification to the 2017 French medical-certificate law, with Paris as the treated city, 2016 as the reference year, and the analysis window 2013--2022 (see Supplementary Policy Analysis for full sample definitions and robustness  \cite{AngristPischke2009,baker2025difference,Callaway2022Did}).
The outcomes are the mean Relative Fat Mass (RFM) and the 10th and 25th percentiles of RFM (lower tail of the distribution).
Although the Milan regulation specified a BMI threshold, we use RFM because weight is unobserved for most models (see Methods: RFM).
Let $c$ index cities, $t$ index calendar years, and $q \in \{\text{mean}, 10, 25\}$ index the outcome (mean, 10th percentile, or 25th percentile of the RFM distribution).
We define a treatment indicator $\mathbb{1}\{c = \text{Milan}\}$ and a post-policy indicator $\mathbb{1}\{t \ge 2006\}$.
The standard TWFE DiD model is,
\begin{equation}
Y_{ctq}
=
\lambda_{cq}
+
\gamma_{tq}
+
\beta_q \,
\mathbb{1}\{c = \text{Milan}\}
\mathbb{1}\{t \ge 2006\}
+
\varepsilon_{ctq},
\label{eq:did_twfe}
\end{equation}
where $Y_{ctq}$ is the outcome for city $c$ in year $t$, $\lambda_{cq}$ and $\gamma_{tq}$ are city and year fixed effects, respectively, and $\beta_q$ captures the average post-policy effect of the Milan regulation on outcome $q$ under the usual parallel trends assumption.
To allow for dynamic treatment effects and to assess pre-policy trends, we estimate a two-way fixed-effects (TWFE) event-study specification,
\begin{equation}
Y_{ctq}
=
\alpha_q
+
\sum_{\tau \neq 2005}
\beta_{q\tau}\,
\mathbb{1}\{c = \text{Milan}\}\,\mathbb{1}\{t = \tau\}
+
\lambda_{cq}
+
\gamma_{tq}
+
\varepsilon_{ctq},
\label{eq:eventstudy}
\end{equation}
where $Y_{ctq}$ denotes the mean RFM or the $q$th percentile of the RFM distribution in city $c$ and year $t$, $\beta_{q\tau}$ is the effect of the Milan regulation in year $\tau$ relative to the reference pre-policy year 2005, $\lambda_{cq}$ are city fixed effects, $\gamma_{tq}$ are year fixed effects, and $\varepsilon_{ctq}$ is an idiosyncratic error term.
The interaction terms $\mathbb{1}\{c = \text{Milan}\}\,\mathbb{1}\{t = \tau\}$ form a saturated set of Milan-by-year indicators, so that the event-study coefficients $\beta_{q\tau}$ trace out the dynamic treatment profile for each outcome $q$.
We estimate Eq.~\eqref{eq:eventstudy} using weighted least squares (WLS), with analytic weights equal to the number of individual records in each city-year-outcome cell, so that estimates can be interpreted as population-weighted average treatment effects \cite{AngristPischke2009,baker2025difference,Callaway2022Did}.
Inference is based on heteroskedasticity-consistent HC1 standard errors: we start from the White (HC0) sandwich estimator and apply a small-sample degrees-of-freedom correction \cite{White1980,MacKinnonWhite1985,MacKinnon2011}.
HC1 is widely used in applied econometrics and simulation evidence shows that, relative to HC0, it typically improves finite-sample size control \cite{MacKinnonWhite1985,MacKinnon2011}.
Given the small number of clusters ($G=5$ cities), we report heteroskedasticity-robust rather than cluster-robust standard errors, as asymptotic cluster inference is unreliable with few clusters \cite{Cameron2015}.
To assess the plausibility of the DiD identifying assumption of parallel pre-policy trends, we implement a standard pre-trend test in the event-study framework \cite{Luedicke2022,RiverosGavilanes2023}.
Specifically, for each outcome $q$ we test the joint null hypothesis that all pre-policy Milan-by-year coefficients are zero,
\begin{equation}
H_0:\ \beta_{q\tau} = 0 \quad \text{for all } \tau < 2006,\ \tau \neq 2005,
\end{equation}
using a Wald test constructed from the HC1 covariance matrix of $\hat{\beta}_{q\tau}$.
Let $\hat{\boldsymbol{\beta}}_q^{\text{pre}}$ denote the vector collecting all pre-policy coefficients $\{\hat{\beta}_{q\tau} : \tau < 2006,\ \tau \neq 2005\}$ and let $\widehat{\mathrm{Var}}(\hat{\boldsymbol{\beta}}_q^{\text{pre}})$ be the corresponding HC1 covariance submatrix.
The Wald statistic is
\begin{equation}
W_q
=
\hat{\boldsymbol{\beta}}_q^{\text{pre}\,\prime}
\left[
\widehat{\mathrm{Var}}(\hat{\boldsymbol{\beta}}_q^{\text{pre}})
\right]^{-1}
\hat{\boldsymbol{\beta}}_q^{\text{pre}},
\end{equation}
which under $H_0$ is asymptotically distributed as a chi-squared random variable with $K_{\text{pre}}$ degrees of freedom, where $K_{\text{pre}}$ is the number of pre-policy coefficients tested \cite{Luedicke2022,RiverosGavilanes2023}.
Following recent guidance, we use the pre-trend Wald test as a diagnostic tool rather than as a mechanical model-selection rule, because conditioning post-treatment inference on prior pre-trend tests can lead to distorted coverage and $p$-values \cite{Roth2022Pretrends}.
All models are estimated in Python using the \texttt{WLS} estimator from the \texttt{statsmodels} package, with city and year fixed effects included as sets of indicator variables, analytic weights equal to the number of underlying observations in each city-year cell, and HC1 heteroskedasticity-consistent standard errors \cite{MacKinnonWhite1985,MacKinnon2011}.

\subsection*{Odds ratio and intersectionality analysis}

To quantify the association between ethnicity and plus-size status, we compute odds ratios comparing non-White to White models.
We classify models as plus-size if their converted US dress size is $\geq 12$, following industry standards~\cite{VogueBusiness_SI_SS2024}.
All non-White categories are collapsed into a single group due to sparse plus-size counts in individual categories.
For each year $t$, let $a_t$, $b_t$, $c_t$, $d_t$ denote the counts of plus-size non-White, non-plus-size non-White, plus-size White, and non-plus-size White models, respectively.
We apply a continuity correction of $0.5$ to all cells to handle sparse counts.
The odds ratio is
\begin{equation}
\mathrm{OR}_t = \frac{(a_t + 0.5)(d_t + 0.5)}{(b_t + 0.5)(c_t + 0.5)}.
\end{equation}
Confidence intervals are computed on the log scale using the standard error
\begin{equation}
\mathrm{SE}(\log \mathrm{OR}_t) = \sqrt{\frac{1}{a_t+0.5} + \frac{1}{b_t+0.5} + \frac{1}{c_t+0.5} + \frac{1}{d_t+0.5}},
\end{equation}
with 95\% CI given by $\exp(\log \mathrm{OR}_t \pm 1.96 \times \mathrm{SE})$.
Statistical significance is assessed using a two-sample $z$-test for proportions.
Let $p_{\text{nw}}$ and $p_{\text{w}}$ denote the plus-size proportions for non-White and White models, with sample sizes $n_{\text{nw}}$ and $n_{\text{w}}$.
Under the null hypothesis of equal proportions, the pooled proportion is $\hat{p} = (a_t + c_t)/(n_{\text{nw}} + n_{\text{w}})$, and the test statistic is
\begin{equation}
z = \frac{p_{\text{nw}} - p_{\text{w}}}{\sqrt{\hat{p}(1-\hat{p})(n_{\text{nw}}^{-1} + n_{\text{w}}^{-1})}},
\end{equation}
with two-sided $p$-values computed from the standard normal distribution.
Yearly odds ratios, confidence intervals, and $p$-values are reported in Table~\ref{tab:intersectional};
sample sizes and plus-size proportions by ethnicity are reported in Table~\ref{tab:intersectional_prop}.

\section*{Data availability}
Model and work-level datasets are available in the Zenodo repository `Cultural evolution of human beauty standards' \href{https://zenodo.org/records/17638160}{https://zenodo.org/records/17638160}.
The general-population benchmark dataset of the US National Health and Nutrition Examination Survey (NHANES), is available at \href{https://www.cdc.gov/nchs/nhanes/}{https://www.cdc.gov/nchs/nhanes/}.

\section*{Code availability}

Code is available at GitHub (\href{https://github.com/LCB0B/evolution-beautystd}{https://github.com/LCB0B/evolution-beautystd})

}

\section*{Competing interests}
The authors declare no competing interests.

\newpage
\section*{Supplementary Figures}

\newpage

\newpage

\newpage

\newpage

\clearpage
\newpage

\clearpage
\newpage

\begin{table}[ht]
\centering
\setlength{\tabcolsep}{6pt}
\renewcommand{\arraystretch}{0.95}
% Start striping from the first data row (row 2), alternate gray/white
\rowcolors{2}{gray!10}{white} % or {rowgray}{white}
\begin{tabular}{lrrrrrrr}
\toprule\toprule
Year & Coefficient & SE & CI (low) & CI (high) & $N_{\text{control}}$ & $N_{\text{treated}}$ & $p$ \\
\midrule
\multicolumn{8}{l}{\textit{\textbf{Outcome: Mean}}} \\
2002 & -0.217 & 0.145 & -0.502 & 0.068 & 4,889 & 542 & 0.148 \\
2003 & -0.111 & 0.116 & -0.339 & 0.117 & 5,812 & 615 & 0.348 \\
2004 & -0.132 & 0.140 & -0.407 & 0.142 & 6,984 & 529 & 0.354 \\
2005 & 0.000 & 0.000 & 0.000 & 0.000 & 8,273 & 655 & - \\
2006 & 0.522 & 0.045 & 0.435 & 0.609 & 9,294 & 699 & $2.0 \times 10^{-11}$ \\
2007 & 0.712 & 0.043 & 0.628 & 0.795 & 10,441 & 621 & $1.1 \times 10^{-14}$ \\
2008 & 1.133 & 0.055 & 1.024 & 1.241 & 10,395 & 665 & $1.1 \times 10^{-16}$ \\
2009 & 0.816 & 0.082 & 0.654 & 0.977 & 11,607 & 784 & $5.7 \times 10^{-10}$ \\
2010 & 0.891 & 0.093 & 0.708 & 1.074 & 13,322 & 1,257 & $1.2 \times 10^{-9}$ \\
\midrule
\multicolumn{8}{l}{\textit{\textbf{Outcome: Q10}}} \\
2002 & -0.057 & 0.097 & -0.248 & 0.133 & 4,889 & 542 & 0.561 \\
2003 & -0.003 & 0.084 & -0.168 & 0.162 & 5,812 & 615 & 0.972 \\
2004 & -0.169 & 0.085 & -0.335 & -0.002 & 6,984 & 529 & 0.058 \\
2005 & 0.000 & 0.000 & 0.000 & 0.000 & 8,273 & 655 & - \\
2006 & 1.352 & 0.052 & 1.250 & 1.455 & 9,294 & 699 & $5.1 \times 10^{-19}$ \\
2007 & 1.364 & 0.056 & 1.255 & 1.474 & 10,441 & 621 & $1.9 \times 10^{-18}$ \\
2008 & 1.330 & 0.057 & 1.219 & 1.442 & 10,395 & 665 & $5.1 \times 10^{-18}$ \\
2009 & 1.105 & 0.110 & 0.889 & 1.321 & 11,607 & 784 & $4.7 \times 10^{-10}$ \\
2010 & 0.993 & 0.071 & 0.853 & 1.132 & 13,322 & 1,257 & $5.2 \times 10^{-13}$ \\
\midrule
\multicolumn{8}{l}{\textit{\textbf{Outcome: Q25}}} \\
2002 & 0.046 & 0.132 & -0.213 & 0.305 & 4,889 & 542 & 0.731 \\
2003 & -0.138 & 0.135 & -0.403 & 0.127 & 5,812 & 615 & 0.317 \\
2004 & -0.284 & 0.160 & -0.599 & 0.030 & 6,984 & 529 & 0.089 \\
2005 & 0.000 & 0.000 & 0.000 & 0.000 & 8,273 & 655 & - \\
2006 & 1.325 & 0.055 & 1.217 & 1.433 & 9,294 & 699 & $2.6 \times 10^{-18}$ \\
2007 & 1.303 & 0.062 & 1.181 & 1.424 & 10,441 & 621 & $5.6 \times 10^{-17}$ \\
2008 & 1.306 & 0.071 & 1.167 & 1.445 & 10,395 & 665 & $1.2 \times 10^{-15}$ \\
2009 & 0.959 & 0.073 & 0.815 & 1.102 & 11,607 & 784 & $2.0 \times 10^{-12}$ \\
2010 & 1.262 & 0.114 & 1.039 & 1.485 & 13,322 & 1,257 & $6.2 \times 10^{-11}$ \\
\bottomrule
\end{tabular}
\caption{\textbf{Event Study Result for Milan (2002-2010)}
For each year and outcome (Mean, Q10, Q25), the table reports the estimated coefficient, standard error (SE), 95\% confidence interval (CI), the number of observations in the control group ($N_{\text{control}}$) and treated group ($N_{\text{treated}}$), and the $p$-value.
The results indicate a significant positive effect in the treated group relative to the control group starting in 2006. We formally test the validity of the research design using a Wald test on the joint significance of the pre-treatment coefficients (2002--2005);
the parallel trends assumption cannot be rejected for any outcome: Mean ($\chi^2=3.53, p=0.317$), Q10 ($\chi^2=4.41, p=0.220$), and Q25 ($\chi^2=4.26, p=0.235$).}
\label{tab:milan_main_results}
\end{table}

\clearpage
\newpage
\begin{table}[ht]
\centering
\setlength{\tabcolsep}{6pt}
\renewcommand{\arraystretch}{0.95}
\rowcolors{2}{gray!10}{white} % or {rowgray}{white}
\begin{tabular}{lrrrrrrr}
\toprule\toprule
Year & Coefficient & SE & CI (low) & CI (high) & $N_{\text{control}}$ & $N_{\text{treated}}$ & $p$ \\
\midrule
\multicolumn{8}{l}{\textit{\textbf{Outcome: Mean}}} \\
2013 & 0.072 & 0.089 & -0.102 & 0.247 & 7,547 & 2,649 & 0.420 \\
2014 & -0.002 & 0.075 & -0.148 & 0.145 & 7,374 & 3,330 & 0.984 \\
2015 & 0.023 & 0.060 & -0.095 & 0.141 & 7,389 & 3,492 & 0.706 \\
2016 & 0.000 & 0.000 & 0.000 & 0.000 & 5,426 & 2,436 & - \\
2017 & 0.110 & 0.082 & -0.052 & 0.271 & 3,627 & 1,531 & 0.190 \\
2018 & -0.075 & 0.067 & -0.207 & 0.056 & 2,470 & 1,094 & 0.268 \\
2019 & -0.140 & 0.056 & -0.249 & -0.031 & 6,494 & 4,344 & 0.016 \\
2020 & -0.024 & 0.053 & -0.129 & 0.080 & 5,605 & 3,348 & 0.653 \\
2021 & -0.122 & 0.149 & -0.413 & 0.169 & 540 & 423 & 0.416 \\
2022 & -0.171 & 0.256 & -0.672 & 0.331 & 2,831 & 1,602 & 0.508 \\
\midrule
\multicolumn{8}{l}{\textit{\textbf{Outcome: Q10}}} \\
2013 & 0.007 & 0.166 & -0.319 & 0.333 & 7,547 & 2,649 & 0.966 \\
2014 & -0.269 & 0.104 & -0.473 & -0.064 & 7,374 & 3,330 & 0.014 \\
2015 & -0.194 & 0.097 & -0.384 & -0.004 & 7,389 & 3,492 & 0.052 \\
2016 & 0.000 & 0.000 & 0.000 & 0.000 & 5,426 & 2,436 & - \\
2017 & -0.381 & 0.137 & -0.649 & -0.113 & 3,627 & 1,531 & 0.008 \\
2018 & -0.088 & 0.138 & -0.358 & 0.182 & 2,470 & 1,094 & 0.526 \\
2019 & -0.189 & 0.067 & -0.320 & -0.057 & 6,494 & 4,344 & 0.008 \\
2020 & -0.208 & 0.067 & -0.340 & -0.076 & 5,605 & 3,348 & 0.004 \\
2021 & 0.044 & 0.138 & -0.226 & 0.314 & 540 & 423 & 0.749 \\
2022 & -0.218 & 0.069 & -0.353 & -0.083 & 2,831 & 1,602 & 0.003 \\
\midrule
\multicolumn{8}{l}{\textit{\textbf{Outcome: Q25}}} \\
2013 & 0.102 & 0.105 & -0.105 & 0.308 & 7,547 & 2,649 & 0.340 \\
2014 & -0.028 & 0.059 & -0.144 & 0.088 & 7,374 & 3,330 & 0.643 \\
2015 & -0.043 & 0.073 & -0.185 & 0.100 & 7,389 & 3,492 & 0.561 \\
2016 & 0.000 & 0.000 & 0.000 & 0.000 & 5,426 & 2,436 & - \\
2017 & 0.043 & 0.070 & -0.093 & 0.180 & 3,627 & 1,531 & 0.537 \\
2018 & -0.003 & 0.052 & -0.106 & 0.099 & 2,470 & 1,094 & 0.948 \\
2019 & -0.194 & 0.112 & -0.415 & 0.026 & 6,494 & 4,344 & 0.090 \\
2020 & -0.069 & 0.072 & -0.210 & 0.071 & 5,605 & 3,348 & 0.339 \\
2021 & 0.342 & 0.154 & 0.041 & 0.643 & 540 & 423 & 0.031 \\
2022 & -0.206 & 0.055 & -0.315 & -0.098 & 2,831 & 1,602 & $5.7 \times 10^{-4}$ \\
\bottomrule
\end{tabular}
\caption{\textbf{Event Study Results for Paris (2013-2022).}
For each year and outcome (Mean, Q10, Q25), the table reports the estimated coefficient, standard error (SE), 95\% confidence interval (CI), the number of observations in the control group ($N_{\text{control}}$) and treated group ($N_{\text{treated}}$), and the $p$-value.
We formally test the validity of the research design using a Wald test on the joint significance of the pre-treatment coefficients;
the parallel trends assumption cannot be rejected for Mean ($\chi^2=0.94, p=0.817$) and Q25 ($\chi^2=2.13, p=0.546$), but shows evidence of differential pre-trends for Q10 ($\chi^2=8.44, p=0.038$).}
\label{tab:paris_main_results}
\end{table}

\clearpage
\newpage

\begin{table}[ht]
\centering
\setlength{\tabcolsep}{6pt}
\renewcommand{\arraystretch}{1.1}
\rowcolors{2}{gray!10}{white} % or {rowgray}{white}
\begin{tabular}{crrrrrr}
\toprule\toprule
Year& $n$ & Odds ratio & OR CI (low) & OR CI (high) & $z$ & $p$ \\
\midrule\midrule
2017 & 31,810 & 2.6 & 1.5 & 4.4 & 3.5 & 0.00046 \\
2018 & 32,615 & 1.9 & 1.1 & 3.1 & 2.5 & 0.012 \\
2019 & 39,137 & 2.3 & 1.5 & 3.6 & 3.7 & 0.00019 \\
2020 & 27,784 & 3.7 & 2.5 & 5.5 & 7.0 & 1.9e-12 \\
2021 & 18,592 & 6.9 & 4.9 & 9.7 & 13 & 5.7e-37 \\
2022 & 24,794 & 4.4 & 3.3 & 5.8 & 11 & 2.1e-28  \\
2023 & 25,326 & 3.4 & 2.5 & 4.6 & 8.1 & 5.1e-16\\
2024 & 21,448 & 4.1 & 2.9 & 5.8 & 8.7 & 3.4e-18 \\
\midrule
2017-2024 & 221,506 & \textbf{4.5} & 3.9 & 5.1 & 25.1 & 1.3e-54 \\ 
\bottomrule
\end{tabular}
\caption{\textbf{Odds ratio of plus-size representation by ethnicity, 2017-2024.}
For each year, the table reports the total number of models ($n$), the odds ratio (or) (non-White vs.\ White), its 95\% confidence interval (CI), the corresponding $z$-statistic, and $p$-value.
Since 2017, non-White models are consistently more likely to be plus-size, with odds ratios increasing sharply after 2020 and remaining highly significant throughout the period.
The value in bold is the value reported in the abstract.
}
\label{tab:intersectional}
\end{table}

\clearpage
\newpage

\begin{table}[ht]
\centering
\setlength{\tabcolsep}{6pt}
\renewcommand{\arraystretch}{1.1}
\rowcolors{2}{gray!10}{white}
\begin{tabular}{lrrrrrr}
\toprule\toprule
Year &
$n_{\text{Non-White}}$ &
$n_{\text{White}}$ &
$n$ &
\% plus Non-White &
\% plus White \\
\midrule\midrule
2017 & 7983 & 23827 & 31810 & 0.301 & 0.118  \\
2018 & 9841 & 22774 & 32615 & 0.274 & 0.145  \\
2019 & 13205 & 25932 & 39137 & 0.310 & 0.135  \\
2020 & 9889 & 17895 & 27784 & 0.769 & 0.207  \\
2021 & 7355 & 11237 & 18592 & 2.379 & 0.347  \\
2022 & 10680 & 14114 & 24794 & 1.882 & 0.432  \\
2023 & 11047 & 14279 & 25326 & 1.276 & 0.378  \\
2024 & 9338 & 12110 & 21448 & 1.424 & 0.347  \\
\midrule
2017--2024 & 79338 & 142168 & 221506 & 0.735 & 0.152 \\
\bottomrule
\end{tabular}
\caption{\textbf{Sample size and proportion of plus-size representation by ethnicity, 2017-2024.}
For each year, the table reports the total number of models ($n$), Non-White models ($n_{\text{Non-White}}$), White models ($n_{\text{White}}$), the share of plus-size models in the two groups of models (\% plus Non-White,\% plus White).}
\label{tab:intersectional_prop}
\end{table}

% -----------------------------------------------------------------------------
% PART 2: SUPPLEMENTARY MATERIAL
% -----------------------------------------------------------------------------
\clearpage
\newpage

% Reset Title and Author for Supplementary
\section*{Supplementary: Cultural Evolution of Human Beauty Standards}

\tableofcontents
\newpage

% FIX PART 2: Turn ON list-of-figures recording for Supplementary & Reset Counters
\captionsetup[figure]{list=yes}
\captionsetup[table]{list=yes}
\setcounter{figure}{0}
\renewcommand{\thefigure}{S\arabic{figure}}
\setcounter{table}{0}
\renewcommand{\thetable}{S\arabic{table}}

\listoffigures
\newpage

\section{Data}
\label{sec:model_data}

\begin{figure}[h!]
  \centering
  \includegraphics[width=\textwidth]{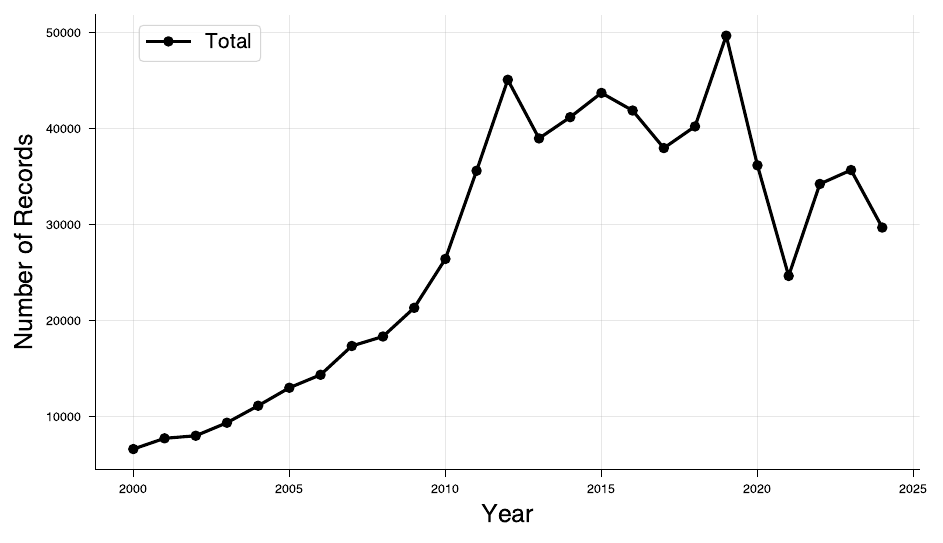}
  \caption{Number of records per year (2000-2024) in the work dataset.}
  \label{fig:si_record_count}
\end{figure}

\begin{figure}[h!]
  \centering
  \includegraphics[width=\textwidth]{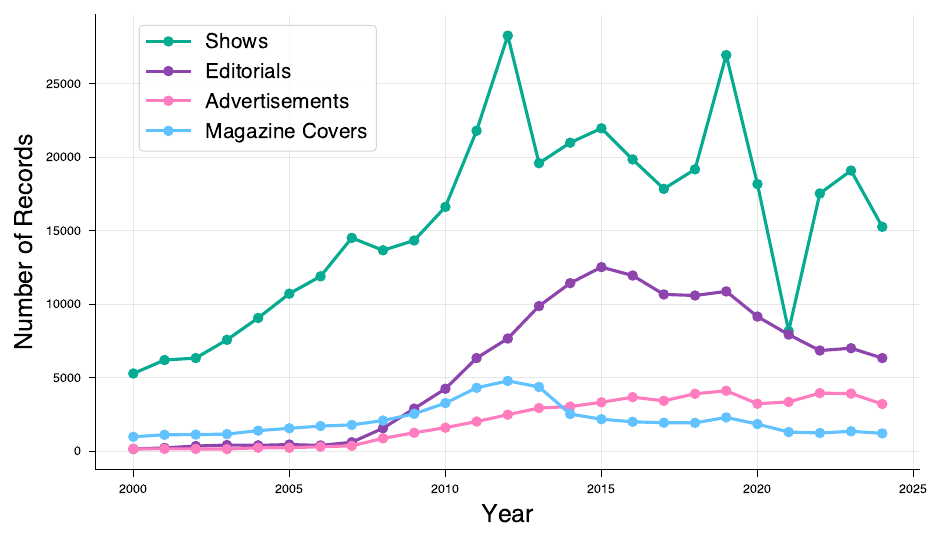}
  \caption{Number of records per year (2000-2024) in the work dataset by work categories.}
  \label{fig:si_record_count_categories}
\end{figure}

\begin{figure}[h!]
  \centering
  \includegraphics[width=\textwidth]{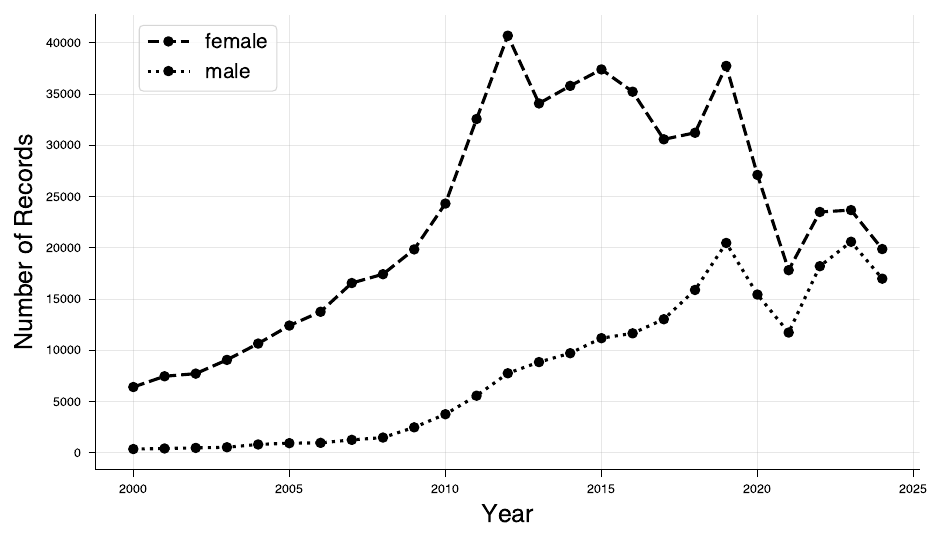}
  \caption{Number of records per year (2000-2024) in the dataset by gender.}
  \label{fig:si_record_count_gender}
\end{figure}

We compiled a dataset from professional pages on \url{https://models.com} and \url{https://www.fashionmodeldirectory.com}.
These platforms aggregate portfolio data submitted by modeling agencies and verified by industry professionals;
profile attributes are typically self-reported by models or provided by their representing agencies.
After de-duplication at the name-agency-URL level, the corpus comprises 793{,}199 records.
Available attributes include anthropometrics (height, bust, waist, hips, dress size, shoe size), identity fields (name, nationality; when available, country of origin, age, weight, gender, ethnicity), Instagram handle, and profile photographs.
Figure \ref{fig:si_record_count} shows the count of records over time. Figure \ref{fig:si_record_count_categories} shows the distribution of work categories over time.
Figure \ref{fig:si_record_count_gender} shows the evolution of the number of records per gender per year, the dataset contains 1{,}753 male models and 13{,}430 female models.

\subsection{Origin and phenotypic classification}

\subsubsection{Regional classification}
\label{si:worldregions}

\begin{figure}[p]
  \vspace*{-2.5cm}
  \centering
  \includegraphics[page=1,width=\textwidth,trim=0 50 0 0,clip]{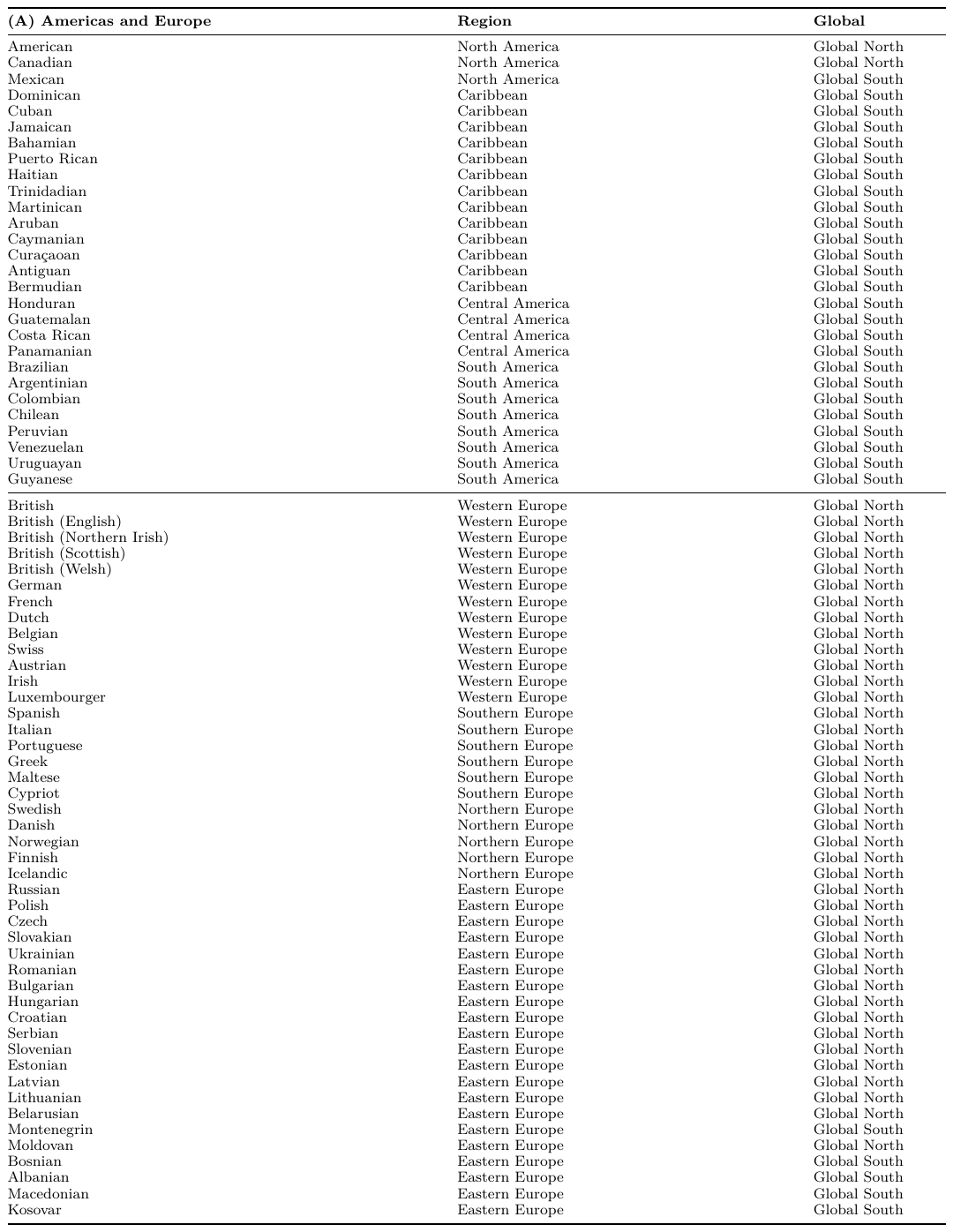}
  \caption[{Mapping of nationalities to regions and Global North/Global South (Panel~A).}]{%
  {Mapping of nationalities to regions and Global North/Global South (Panel~A: Americas and Europe).}
  }
  \label{fig:world_regions_A}
\end{figure}

\begin{figure}[p]
  \vspace*{-2cm}
  \centering
  \includegraphics[page=2,width=\textwidth,trim=0 50 0 0,clip]{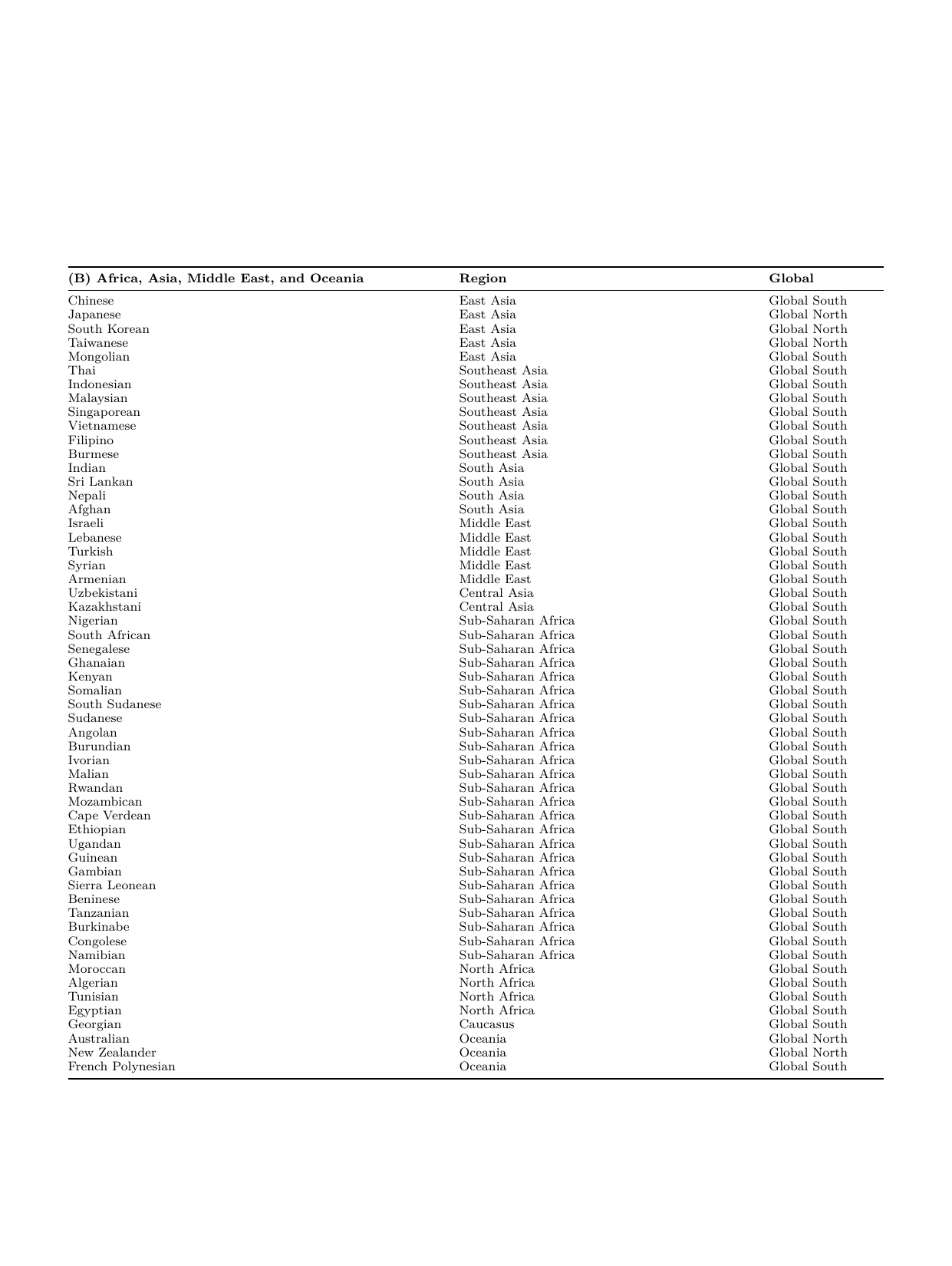}
  \caption[{Mapping of nationalities to regions and Global North/Global South (Panel~B).}]{%
  {Mapping of nationalities to regions and Global North/Global South (Panel~B: Africa, Asia, Middle East, and Oceania).}
  }
  \label{fig:world_regions_B}
\end{figure}

Figures \ref{fig:world_regions_A} and \ref{fig:world_regions_B} illustrate the mapping of nationalities across broader world regions and the categories of the \emph{Global North} and \emph{Global South}.
This classification follows common practice in development studies and international relations, where the terms \emph{Global North} and \emph{Global South} are used as shorthand to describe patterns of socioeconomic development and geopolitical positioning \cite{brandt_report1980, lees2021brandt_line, prashad_metanalysis}.

% Panel A

% Panel B

\clearpage
\newpage

\subsubsection{Gender and ethnicity attribution}
\label{si:gender_ethnicity}

\begin{figure}[h]
  \centering
  \includegraphics[page=2,width=\textwidth]{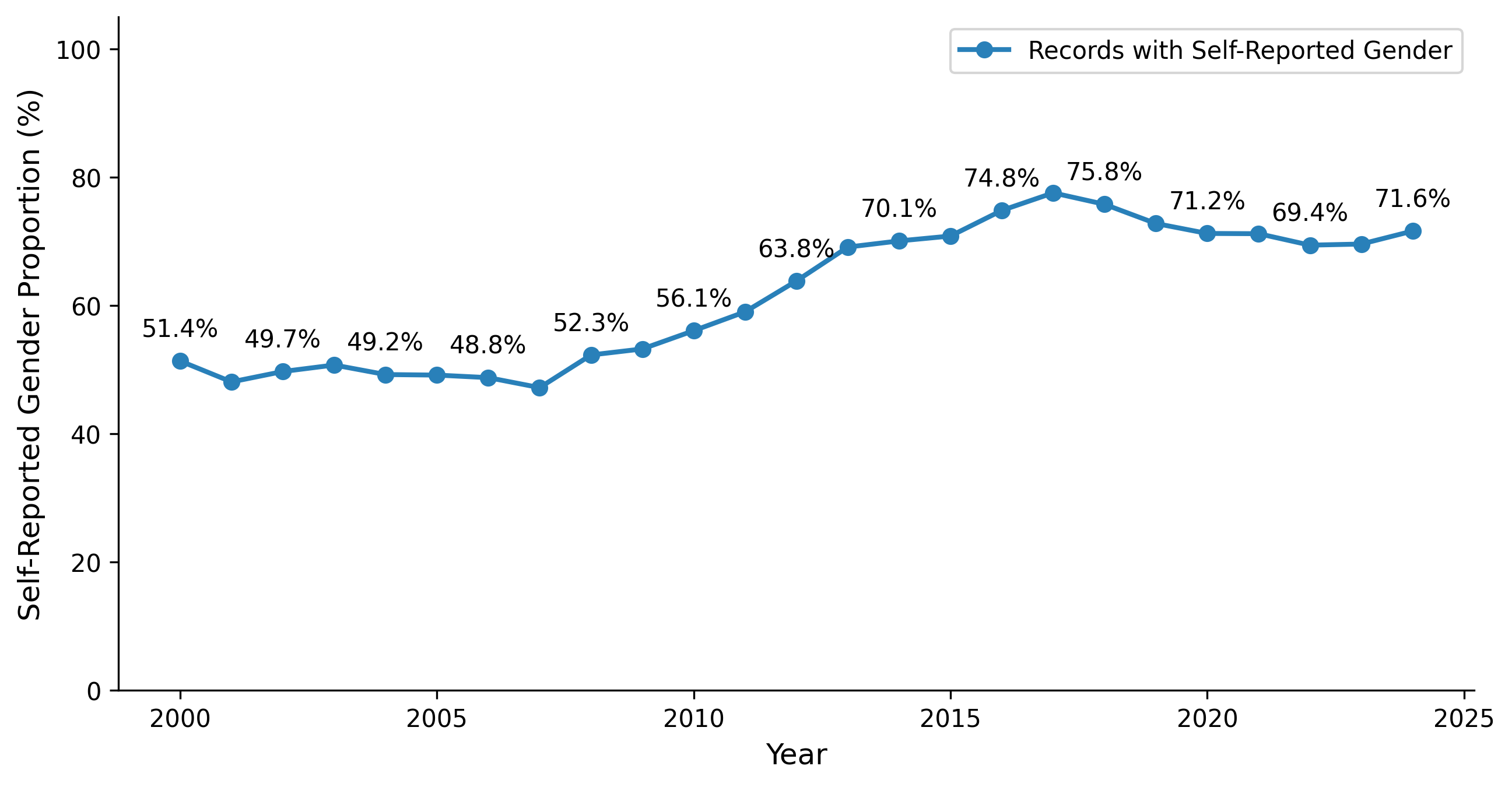}
  \caption[Self-reported gender proportion]{Self-reported gender coverage over time for 
records.
  }
  \label{fig:si-gender_coverage_evolution}
\end{figure}

\begin{figure}[h]
  \centering
  \includegraphics[page=2,width=\textwidth]{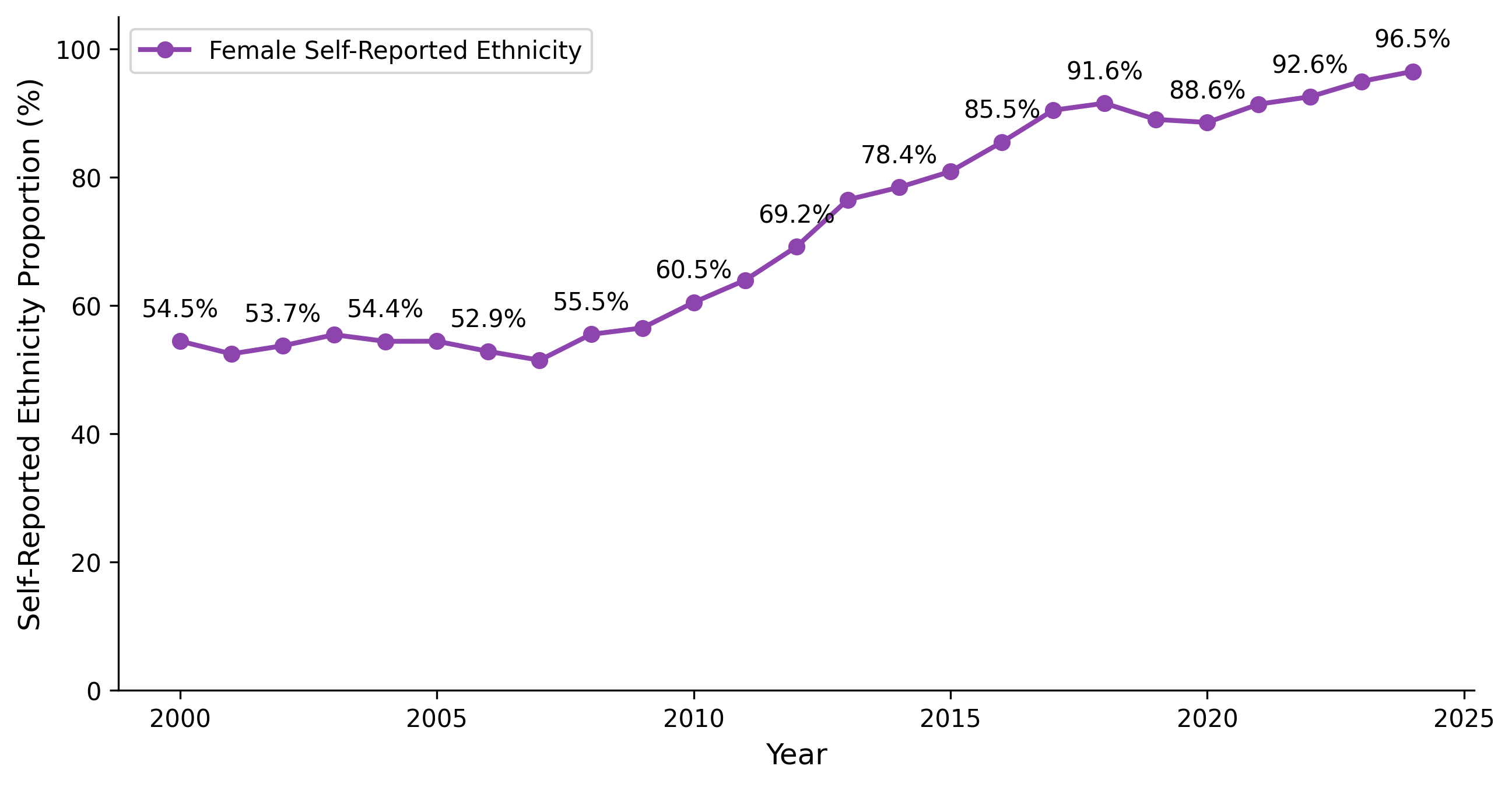}
  \caption[Self-reported Ethnicity proportion]{Self-reported ethnicity coverage over time for female record.
}
  \label{fig:si-female_ethnicity_coverage_evolution}
\end{figure}

The data coverage on self-reported gender and ethnicity is partial, over the study period (2000-2024) 64.3\% of the record have a self reported gender and 66.2\% have a self-reported ethnicity (77.0\% self reported ethnicity for female for female).
Figure \ref{fig:si-gender_coverage_evolution} shows the evolution of the proportion of records with self-reported gender.
Figure \ref{fig:si-female_ethnicity_coverage_evolution} shows the evolution of the proportion of records with self-reported ethnicity for female models.
For the remaining records where self-reported attributes are unavailable, we rely on image-based predictions to ensure comprehensive demographic coverage.
We acknowledge that the gender binary classification oversimplifies the complex spectrum of gender identities and may inadvertently exclude or misclassify transgender, non-binary, and gender-nonconforming individuals.
Categorizing models by gender is crucial for analyzing changing beauty standards, as female and male models generally display distinct body measurement characteristics.
To infer gender, we use an image-based classification via the \texttt{FairFace} deep learning model, which predicts perceived gender from facial images \cite{karkkainen2021fairface}.
Race is classified based on visual appearance rather than nationality, as nationality does not necessarily reflect racialized representation in fashion imagery.
This approach allows for a more direct measurement of visible diversity in the modeling industry.
We use the \texttt{FairFace} model \cite{karkkainen2021fairface}, a convolutional neural network trained on a balanced dataset across seven race categories: White, Black, East Asian, South East Asian, Latino, Indian, and Middle Eastern.
Following the authors' implementation guidelines, we employ the \texttt{dlib} face detection model from the \texttt{dlib-models} repository to detect faces in profile images from the model dataset.
Detected faces are cropped and aligned according to the \texttt{FairFace} preprocessing pipeline, ensuring consistency with the model’s training setup.
The processed images are then passed through the model to obtain predicted race categories.
For the main analysis, all non-White categories are collapsed into a single “non-White” group to account for the limited representation of plus-sized models in some racial subcategories.
Models with an “Unknown” classification or with no detected face are excluded from the analysis.
\begin{table}[htb]\centering
\begin{tabular}{p{6cm}lllll}
\hline
\textbf{Validation Task} & \textbf{Precision} & \textbf{Recall} & \textbf{F1 Score} & \textbf{Accuracy} & \textbf{N} \\
\hline
Gender (Man/Woman) & 0.97 & 0.94 & 0.95 & 0.93 & 4,378 \\
Ethnicity (non-White/White) & 0.89 & 0.91 & 0.90 & 0.85 & 4,428 \\
\hline
\end{tabular}
\caption{\textbf{Validation results for gender and ethnicity classification.} The table reports weighted-average Precision, Recall, F1 Score, and unweighted Accuracy, along with the number of observations (N).}
\label{tab:validation_metrics}
\end{table}

We validated the accuracy of the gender and ethnicity attribution procedures by comparing the combined name- and image-based classifications with a subset of models for whom self-reported gender and ethnicity were available (respectively $N = 4{,}378$ and $N = 
4{,}428$). The results, summarized in Table \ref{tab:validation_metrics}, show high agreement between sources.
The gender classification achieves an F1 score of $0.95$, and the ethnicity (White vs.\ non-White) classification reaches a weighted F1 score of $0.90$.
These results show that both attribution procedures perform reliably. The validation for ethnicity aligns with the range reported by the authors of the \texttt{FairFace} model ($75.5\%$ to $97\%$, depending on the ethnic group and dataset)  \cite{karkkainen2021fairface}.
Gender and ethnicity attribution could be further improved by aggregating predictions across multiple images per model and assigning the final label via majority voting across the detected categories.
\subsubsection{Plus-size attribution}
\label{si:plus_size}

The information on models comes from their self-described dress size.
We adopt the standard fashion industry definition of plus-size, defined as models with a U.S. dress size $\geq 12$.
Each model record contains two dress size measures, one in EU sizing and one in U.S. sizing.
The raw size information appears in heterogeneous string formats that may mix conventions across regions, for example:

\begin{itemize}
    \item Single numeric values (e.g., \texttt{38} or \texttt{10}),
    \item Size ranges (e.g., \texttt{36--38} or \texttt{8--10}),
    \item Text strings with non-numeric characters (e.g., ``EU 38'', ``Size 12'').
\end{itemize}

We process these values through two coordinated parsers, one for U.S. sizes and one for EU sizes, that follow a set of deterministic rules:

\begin{enumerate}
    \item Remove all non-numeric characters except hyphens.
    \item Split ranges on the hyphen and convert each element to an integer.
    \item For U.S.-sized entries:
    \begin{itemize}
        \item If all values are $\leq 18$, treat as U.S. sizes and take the highest value, rounding up to the nearest even number (as standard dress sizes are typically even-numbered).
        \item If all values are $>18$, treat as EU sizes and convert using the EU$\to$U.S. mapping below.
        \item If a mix of EU and U.S. values appears (some $\leq 18$, some $>18$), use the lower (U.S.) value.
    \end{itemize}
    \item For EU-sized entries:
    \begin{itemize}
        \item Take the highest EU value, round up to the nearest even number, and convert to the corresponding U.S. size.
    \end{itemize}
\end{enumerate}

The EU$\to$U.S. conversion follows the standard industry mapping shown below:

\begin{center}
\begin{tabular}{c|ccccccccccccc}
\toprule
EU size & 30 & 32 & 34 & 36 & 38 & 40 & 42 & 44 & 46 & 48 & 50 & 52 & 54 \\
U.S. size &  0 &  2 &  4 &  6 &  8 & 10 & 12 & 14 & 16 & 18 & 20 & 22 & 24 \\
\bottomrule
\end{tabular}
\end{center}

To ensure reliability, we retain only observations where the independently parsed EU and U.S. sizes are consistent.
All other cases (including missing or inconsistent conversions) are excluded from the analysis ($N = 173$).
Finally, we define a binary variable, \texttt{plus\_sized}, equal to 1 for models with a U.S. dress size~$\geq 12$ and 0 otherwise.
\subsection{Reference population and body metrics}

\subsubsection{General population data}
\label{si:general_population}
\begin{figure}[h!]
    \centering
    \includegraphics[width=\textwidth]{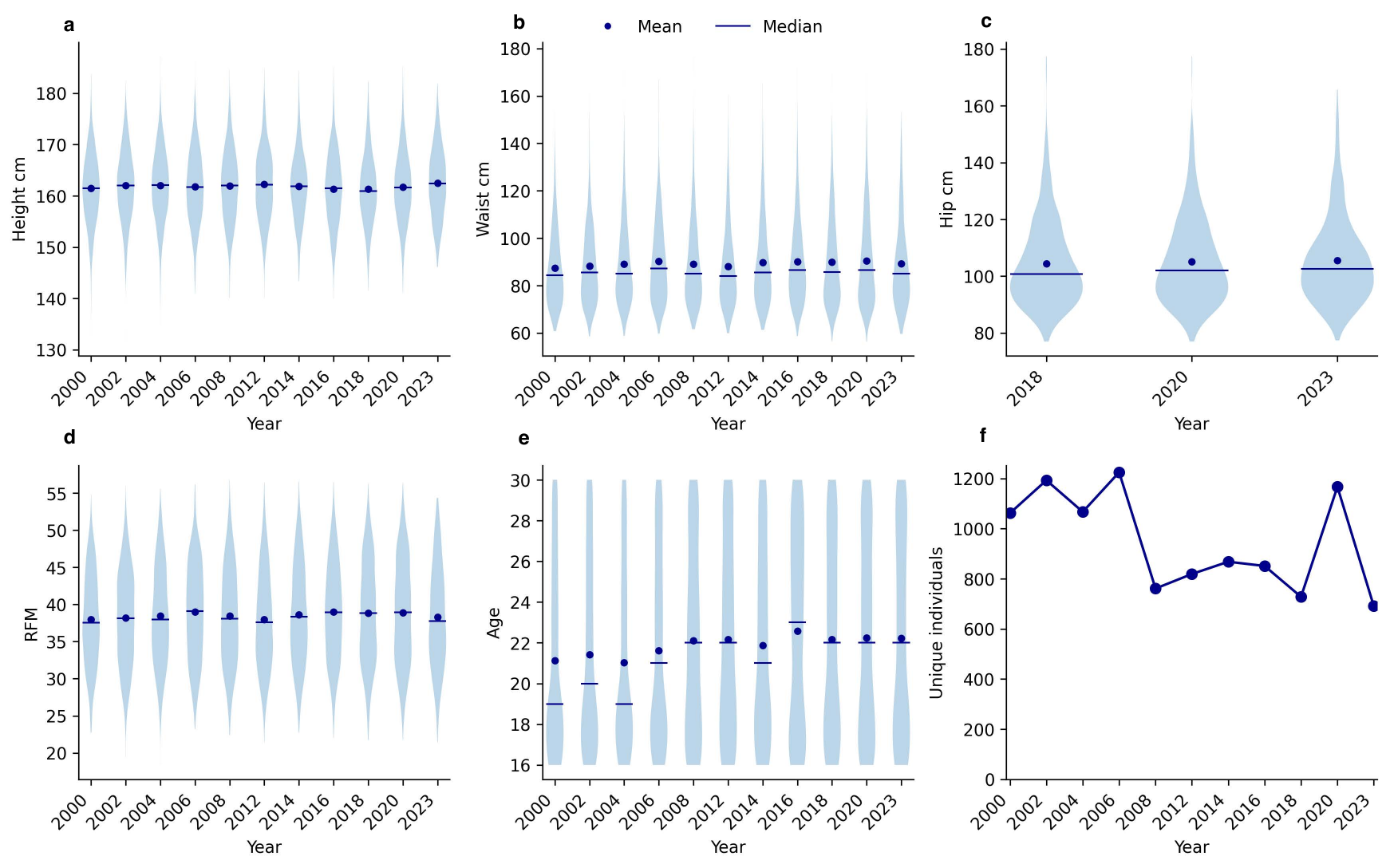}
    \caption[{Anthropometric distributions over time for NHANES females (aged 16-29).}]{%
        \textbf{Anthropometric distributions over time for NHANES females (ages 16-29).}
        Violin plots show the distribution of \textbf{(a)} height, \textbf{(b)} waist circumference, 
        \textbf{(c)} hip circumference, \textbf{(d)} relative fat mass (RFM), and \textbf{(e)} age across years.
        Dots indicate yearly means and horizontal lines mark medians. 
        Panel \textbf{(f)} shows the number of unique individuals contributing data per year.%
    }
    \label{fig:nhanes_distribution}
\end{figure}

To compare body measurements of American models with the general population, we use demographic and anthropometric data from the 2021-2023 cycle of the National Health and Nutrition Examination Survey (NHANES) \cite{NHANES2021_2023}.
NHANES provides a nationally representative sample of the U.S. population, making it suitable for benchmarking model body metrics against general population averages.
To approximate the age range of the models, we restrict the sample to individuals aged 16-29.
Using gender, height, and waist measurements, we compute the Relative Fat Mass (RFM) metric (see \nameref{si:rfm}), which provides a more accurate estimate of body fat percentage than the traditional body mass index (BMI) \cite{woolcott2018relative, Woolcott2020}.
Figure \ref{fig:nhanes_distribution} summarizes the anthropometric variability in the NHANES reference sample across survey years, illustrating stable mean and median measurements for the restricted age group (16-29).

\subsubsection{Relative fat mass}
\label{si:rfm}

Relative Fat Mass (RFM) is a body composition index that estimates an individual's body fat percentage using height, waist circumference, and gender, which was found to better predict whole-body fat percentage among adult individuals than the body mass index (BMI) \cite{woolcott2018relative}.
It more accurately predicts mortality risk and reduces underdiagnosis of obesity \cite{Woolcott2020}.
It is defined as: 
\begin{equation}
\text{RFM} = 64 - 20 \times \left( \frac{\text{height}}{\text{waist}} \right) + 12 \times \text{sex}, \quad \text{where sex} =
\begin{cases}
0 & \text{for men} \\
1 & \text{for women}
\end{cases}
\end{equation}

\section{Statistical estimation}
\label{si:stat}

\subsection{Higher-moment estimators and empirical quantiles}

Let $x_1,\dots,x_n$ denote a sample of $n$ observations, with sample mean
\[
\bar{x}=\frac{1}{n}\sum_{i=1}^n x_i .
\]

We use the unbiased sample variance,
\[
s^2 \;=\; \frac{1}{n-1}\sum_{i=1}^n (x_i-\bar{x})^2 ,
\qquad
s = \sqrt{s^2}.
\]

Skewness is the standardized third central moment, corrected for finite-sample bias:
\[
g_1 \;=\; \frac{n}{(n-1)(n-2)} \sum_{i=1}^n 
\left(\frac{x_i-\bar{x}}{s}\right)^3 .
\]

Excess kurtosis is the standardized fourth central moment, also corrected for
finite-sample bias:
\[
g_2 \;=\;
\frac{n(n+1)}{(n-1)(n-2)(n-3)} 
\sum_{i=1}^n \left(\frac{x_i-\bar{x}}{s}\right)^4
\;-\; \frac{3(n-1)^2}{(n-2)(n-3)} .
\]
By convention, a normal distribution has $g_2=0$.

For distributional summaries we also report empirical quantiles at selected
probabilities $q \in \{0.10,0.25,0.50,0.75,0.90,0.95,0.99\}$, defined by the
order statistics with linear interpolation between sample points.
\subsection{Error bars and uncertainty estimation}

To quantify statistical uncertainty we report error bars for the estimators.
For a sample of size $n$, the standard error of the mean is
\[
\mathrm{SE}(\bar{x}) = \frac{s}{\sqrt{n}} ,
\]
where $s$ is the unbiased sample standard deviation.
The standard error of the sample standard deviation is approximated by
\[
\mathrm{SE}(s) \;\approx\;
\frac{s}{\sqrt{2(n-1)}} ,
\]
which follows from the $\chi^2$ distribution of the sample variance.
For skewness and kurtosis, closed-form standard errors assume Gaussian data 
and are not appropriate in our setting.
Instead, we estimate their uncertainty 
using nonparametric bootstrap resampling: from the original sample, we draw 
$B$ bootstrap replicates with replacement, compute the statistic of interest 
for each replicate, and take the standard deviation of these bootstrap 
replicates as the error estimate.
For quantiles $\hat{Q}(q)$ at probability levels $q \in \{0.10,0.25,0.50,0.75,0.90,0.95,0.99\}$, 
variability likewise depends on the unknown density at the quantile.
We therefore 
use the same bootstrap procedure to obtain error bars.
Unless otherwise noted, error bars in our figures correspond to 
$\pm 1$ bootstrap standard error for skewness, kurtosis, and quantiles, and to 
the analytic formulas above for the mean and standard deviation.
\subsection{Trend analysis}

To assess temporal trends in body measurements across different media types, we employed non-parametric statistical methods that do not assume linearity or normality in the underlying data distribution.
All analyses were conducted on the 25-year time series spanning 2000 to 2024 ($n = 25$ observations per measurement-media combination).
\subsubsection{Mann-Kendall trend test}

The Mann-Kendall test \cite{Mann1945, Kendall1975} was used as the primary method for detecting monotonic trends in the time series data.
This non-parametric test evaluates whether there is a statistically significant trend without making assumptions about the functional form of the relationship.
The Mann-Kendall statistic $S$ is calculated as:

\begin{equation}
S = \sum_{i=1}^{n-1} \sum_{j=i+1}^{n} \text{sgn}(x_j - x_i)
\end{equation}

where $x_i$ and $x_j$ are data values at times $i$ and $j$ respectively ($j > i$), $n$ is the number of data points, and $\text{sgn}$ is the sign function:

\begin{equation}
\text{sgn}(x) = \begin{cases}
+1 & \text{if } x > 0 \\
0 & \text{if } x = 0 \\
-1 & \text{if } x < 0
\end{cases}
\end{equation}

The variance of $S$ under the null hypothesis of no trend is:

\begin{equation}
\text{Var}(S) = \frac{n(n-1)(2n+5) - \sum_{t} t(t-1)(2t+5)}{18}
\end{equation}

where $t$ represents the number of tied values in each tied group.
The normalized test statistic $Z$ is calculated as:

\begin{equation}
Z = \begin{cases}
\frac{S-1}{\sqrt{\text{Var}(S)}} & \text{if } S > 0 \\
0 & \text{if } S = 0 \\
\frac{S+1}{\sqrt{\text{Var}(S)}} & \text{if } S < 0
\end{cases}
\end{equation}

The two-tailed p-value is computed as $p = 2[1 - \Phi(|Z|)]$, where $\Phi$ is the cumulative distribution function of the standard normal distribution.
Kendall's tau ($\tau$), a measure of correlation strength, was calculated as:

\begin{equation}
\tau = \frac{S}{\binom{n}{2}} = \frac{2S}{n(n-1)}
\end{equation}

\subsubsection{Sen's slope estimator}

To quantify the magnitude of detected trends, we employed Sen's slope estimator \cite{Sen1968}, which provides a robust estimate of the rate of change that is insensitive to outliers.
Sen's slope is defined as the median of all pairwise slopes:

\begin{equation}
\text{Sen's slope} = \text{median}\left\{\frac{x_j - x_i}{j - i} : i < j\right\}
\end{equation}

The corresponding intercept was calculated as:

\begin{equation}
\text{intercept} = \text{median}(x_i - \text{slope} \times t_i)
\end{equation}

where $t_i$ represents the time point corresponding to observation $x_i$.
Confidence intervals for Sen's slope were constructed using a rank-based approach.
For a given significance level $\alpha$, the confidence interval bounds are determined by:

\begin{equation}
C_\alpha = \text{round}\left[Z_{1-\alpha/2} \sqrt{\frac{n(n-1)(2n+5)}{18}}\right]
\end{equation}

where $Z_{1-\alpha/2}$ is the $(1-\alpha/2)$ quantile of the standard normal distribution.
\subsubsection{Statistical significance and effect size}

Statistical significance was assessed at $\alpha = 0.05$.
Trends were classified as "increasing," "decreasing," or "no trend" based on the sign of the Mann-Kendall $S$ statistic and the corresponding p-value.
To assess practical significance, we calculated the total change over the study period: $\Delta = \text{Sen's slope} \times \text{time span}$, and the relative change: $\Delta_{\text{rel}} = \frac{\Delta}{\text{baseline median}} \times 100\%$, where the baseline median represents the median value across all time points for each measurement-media combination.
\section{Results}

\subsection{Male sample (2010-2025)}
\label{si:male}

\begin{figure}[h!]
    \centering
    \includegraphics[width=\textwidth]{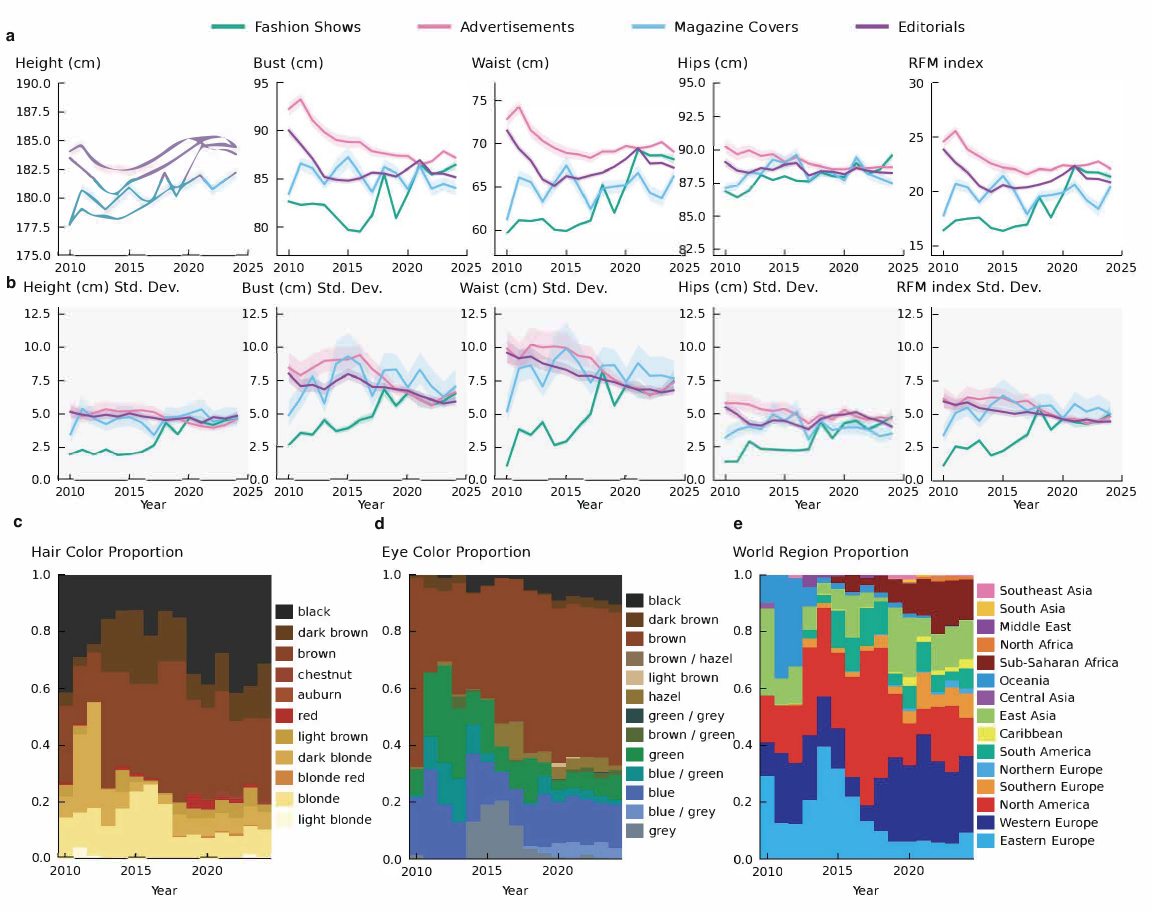}
    \caption[{Evolution of representational diversity for male (2010-2025).}]{%
        \textbf{Evolution of representational diversity for male (2010-2025).}
        \textbf{(a)} Means of height, chest, waist, hips, and RFM by work type (\textcolor{teal}{fashion shows}, \textcolor{softpink}{advertisements}, \textcolor{lightblue}{magazine covers}, and \textcolor{purple}{editorials}).
        Values are in centimeters; shaded areas show confidence intervals.
        \textbf{(b)} Standard deviations remain largely stable, with slight widening after 2020. 
        \textbf{(c)} Stacked shares of hair colors over time.
        \textbf{(d)} Stacked shares of eye colors over time.
        \textbf{(e)} Stacked shares of national origins by world region (see \nameref{si:worldregions}).%
    }
    \label{fig:si-evolution-male}
\end{figure}

Male representation in fashion and media remains limited \cite{mears2010size,jestratijevic2022body}, and this imbalance is reflected in the smaller and more variable male sample (see \nameref{sec:model_data}).
Mean height and body measures (chest, waist, hips) remain stable from 2010 to 2025 across all work types, with no directional trend (Fig.\ref{fig:si-evolution-male}a).
The male physique consistently approximates a balanced 90-70-90 cm contour, with only minor broadening after 2020 (Fig.~\ref{fig:si-evolution-male}b).
Relative fat mass (RFM) remains below population levels but rise slightly over time, suggesting limited relaxation of thinness standards.
Hair and eye color distributions are dominated by darker tones, and regional origins remain concentrated in Western Europe and North America (Fig.~\ref{fig:si-evolution-male}c-e).
These proportions indicate low phenotypic and geographic diversity among male models compared with the female sample.
\begin{figure}[h!]
    \centering
    \includegraphics[width=\textwidth]{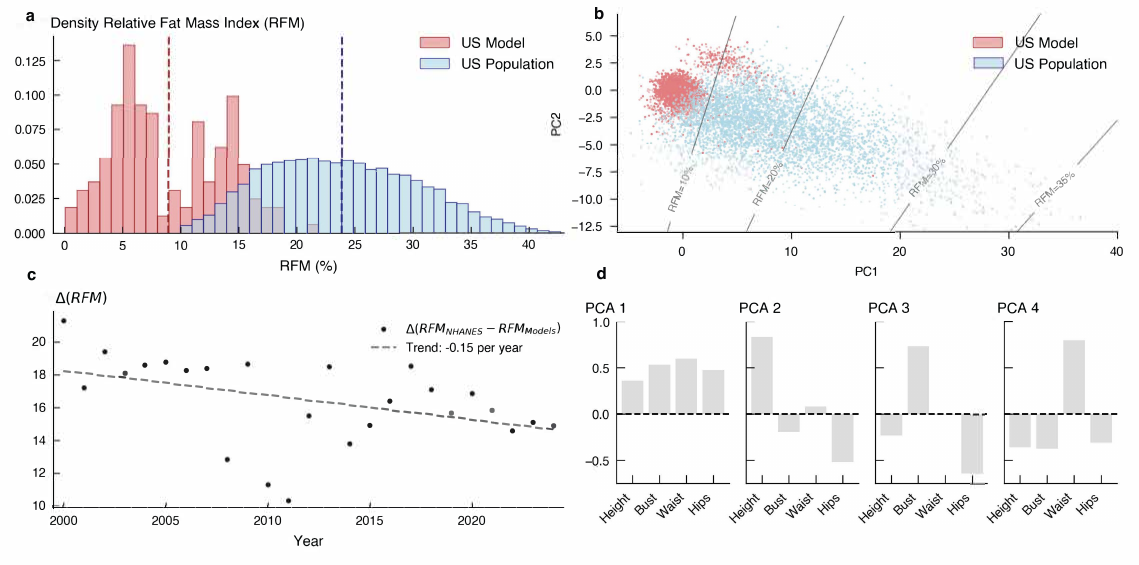}
    \caption[{Distribution shape and population benchmarking for male models.}]{%
        \textbf{Distribution shape and population benchmarking for male models.}
        \textbf{(a)} Density distributions of relative fat mass (RFM) for \textcolor{red}{U.S.-based male models} and \textcolor{blue}{U.S.
        men} aged 16-29 (NHANES 2021-2023), showing limited overlap.
        \textbf{(b)} Principal component analysis (PCA) reveals clear separation between model and population samples, primarily along the adiposity dimension.
        \textbf{(c)} Annual RFM gap ($\Delta$RFM; NHANES - models) with fitted linear trend ($\approx -0.15$ per year) indicates only slight convergence toward the population average.
        \textbf{(d)} PCA loadings show that waist and hips dominate PC1, while height drives PC2.
        }
    \label{fig:si-iqr-male}
\end{figure}

Relative fat mass (RFM) distributions place male models well below the U.S. male population baseline, indicating persistent thinness standards (Fig.~\ref{fig:si-iqr-male}a).
Principal component analysis reveals a clear separation between model and population samples, primarily along adiposity-related axes (Fig.~\ref{fig:si-iqr-male}b).
The annual RFM gap ($\Delta$RFM) shows a modest convergence toward population averages ($\sim-0.15$ per year; Fig.~\ref{fig:si-iqr-male}c), while the principal component loadings (Fig.~\ref{fig:si-iqr-male}d) confirm that waist and hip circumference dominate variation in PC1, with height contributing to PC2.
Overall, male models remain markedly leaner than the general population, but their body measurements exhibit greater variability and overlap with population norms than observed for female models.
This broader spread suggests a comparatively weaker selection pressure on male body types in media and fashion representation.
\begin{table}[h]
\centering
\begin{tabular}{lcccc}
\hline
\textbf{Measurement/Media Type} & \textbf{Sen's Slope} & \textbf{P-value} & \textbf{Kendall's $\tau$} & \textbf{Change (\%)} \\
\hline
\multicolumn{5}{l}{\textit{Height (cm)}} \\
\quad Fashion Shows & 0.025 & 3.62 $\times$ 10$^{-1}$ & 0.13 & 30.3 \\
\quad Advertisements & 0.085 & 1.15 $\times$ 10$^{-7}$ & 0.76 & 66.4 \\
\quad Magazine Covers & 0.013 & 1.83 $\times$ 10$^{-1}$ & 0.19 & 11.0 \\
\quad Editorials & 0.066 & 1.01 $\times$ 10$^{-5}$ & 0.63 & 57.2 \\
\hline
\multicolumn{5}{l}{\textit{Bust Size (cm)}} \\
\quad Fashion Shows & 0.097 & 1.91 $\times$ 10$^{-7}$ & 0.75 & 84.4 \\
\quad Advertisements & 0.154 & 8.13 $\times$ 10$^{-10}$ & 0.88 & 84.1 \\
\quad Magazine Covers & 0.121 & 4.50 
$\times$ 10$^{-10}$ & 0.89 & 82.8 \\
\quad Editorials & 0.140 & 1.38 $\times$ 10$^{-8}$ & 0.81 & 92.1 \\
\hline
\multicolumn{5}{l}{\textit{Waist Size (cm)}} \\
\quad Fashion Shows & 0.079 & 2.91 $\times$ 10$^{-5}$ & 0.60 & 97.2 \\
\quad Advertisements & 0.148 & 8.31 $\times$ 10$^{-7}$ & 0.71 & 76.8 \\
\quad Magazine Covers & 0.099 & 2.59 $\times$ 10$^{-9}$ & 0.85 & 82.3 \\
\quad Editorials & 0.110 & 8.17 $\times$ 10$^{-6}$ & 0.64 & 73.1 \\
\hline
\multicolumn{5}{l}{\textit{Hip Size (cm)}} \\
\quad Fashion Shows & 0.055 & 9.61 $\times$ 10$^{-5}$ & 0.56 & 59.9 \\
\quad Advertisements & 0.123 & 2.38 $\times$ 10$^{-8}$ & 0.80 & 95.9 \\
\quad 
Magazine Covers & 0.076 & 6.85 $\times$ 10$^{-8}$ & 0.77 & 70.1 \\
\quad Editorials & 0.091 & 1.69 $\times$ 10$^{-6}$ & 0.69 & 80.4 \\
\hline
\multicolumn{5}{l}{\textit{Relative Fat Mass}} \\
\quad Fashion Shows & 0.069 & 2.91 $\times$ 10$^{-5}$ & 0.60 & 53.9 \\
\quad Advertisements & 0.101 & 2.38 $\times$ 10$^{-8}$ & 0.80 & 72.4 \\
\quad Magazine Covers & 0.083 & 7.99 $\times$ 10$^{-9}$ & 0.83 & 67.6 \\
\quad Editorials & 0.093 & 8.31 $\times$ 10$^{-7}$ & 0.71 & 68.0 \\
\hline
\end{tabular}
\caption{Mann-Kendall trend test with Sen's slope estimator over 24 years (2000-2024), n=25 observations.
Change (\%) represents relative change from baseline (2000) to final year (2024).}
\label{tab:trends}
\end{table}

\subsection{Temporal trends in female body measurements}
\label{si:temporal_trends_female}

\begin{figure}[h]
    \centering
    \includegraphics[width=\textwidth]{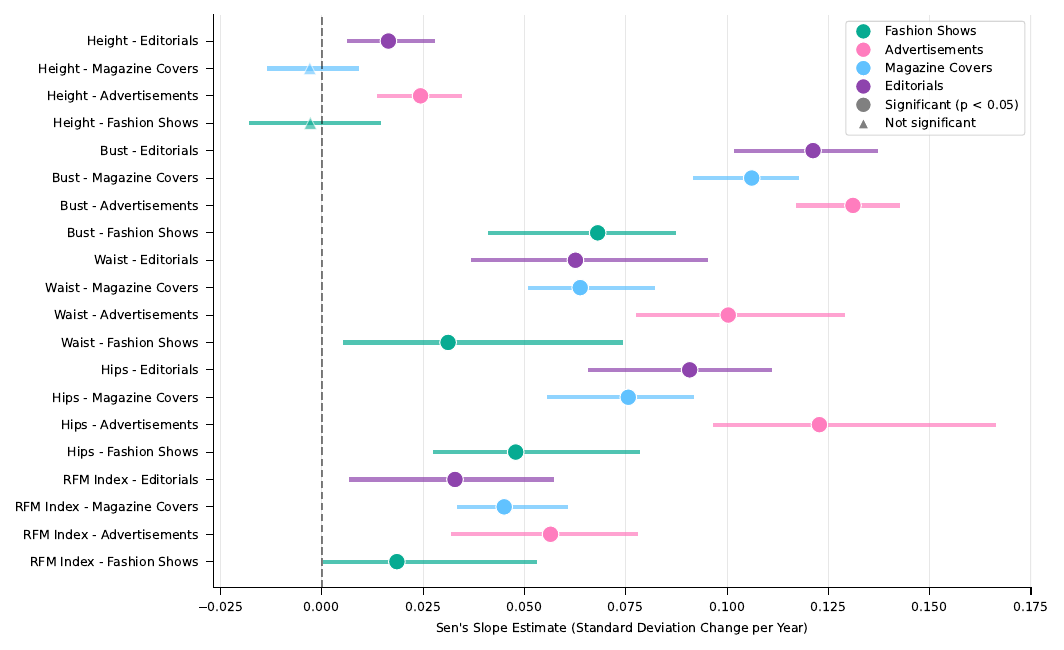}    
    \caption{Standard deviation trends and slope estimates for body measurements (2000-2024).}
    \label{fig:si-forest-plot-std}
\end{figure}

The Mann-Kendall trend analysis reveals statistically significant increases in female body measurements across most media types between 2000 and 2024. Of the 20 measurement-media combinations analyzed, 18 showed significant positive trends ($p < 0.05$), with only height measurements in fashion shows and magazine covers not statistically significant.
Table \ref{tab:trends} summarizes the full results, and Figure \ref{fig:si-forest-plot-std} visualizes the corresponding slope estimates and confidence intervals.
Trend magnitudes differ across measurements and media, with Sen’s slopes spanning from 0.013 for height in magazine covers (non-significant) to 0.154 for bust size in advertisements.
Height shows the most variable pattern across media types. While advertisements and editorials display significant increasing trends (Sen's slopes of 0.085 and 0.066, respectively; both $p < 0.001$), fashion shows and magazine covers show no significant temporal changes ($p = 0.362$ and $p = 0.183$, respectively).
The relative changes over the study period are substantial for the significant trends, with advertisements showing a 66.4\% increase and editorials a 57.2\% rise from baseline values.
All body size measurements (bust, waist, and hip sizes) exhibit consistent and significant increasing trends across all media types.
Bust size shows particularly strong trends, with $p$-values ranging from $1.38 \times 10^{-8}$ to $1.91 \times 10^{-7}$ and Kendall's $\tau$ values between 0.75 and 0.89, indicating robust monotonic increases.
Relative changes range from 82.8\% (magazine covers) to 92.1\% (editorials) over the 24-year period.
Waist size demonstrates similar consistency, with all media types showing significant increases ($p$-values from $2.91 \times 10^{-5}$ to $2.59 \times 10^{-9}$).
Fashion shows record the largest relative change (97.2\%), while editorials showed the smallest but still considerable increase (73.1\%).
Hip size follows the same universal upward trend across media types ($p$-values $9.61 \times 10^{-5}$ - $2.38 \times 10^{-8}$), with the strongest relative change in advertisements (95.9\%) and the weakest in fashion shows (59.9\%).
Relative fat mass measurements show significant increasing trends across all media types, with $p$-values between $2.91 \times 10^{-5}$ and $7.99 \times 10^{-9}$.
The relative changes were more moderate compared to body size measurements, ranging from 53.9\% (fashion shows) to 72.4\% (advertisements), but remained substantial across the two decades.
The strength of detected trends, expressed by Kendall's $\tau$, varied moderately across measurements and media.
The strongest correlations ($\tau > 0.8$) occur for bust size in magazine covers and advertisements ($\tau = 0.89$ and $0.88$, respectively), waist size in magazine covers ($\tau = 0.85$), and RFM in magazine covers ($\tau = 0.83$).
These high $\tau$ values indicate very strong monotonic relationships between time and the measured quantities.
The consistency of significant increases across body dimensions and media types points to a systematic, industry-wide shift in female body representation.
Rather than isolated exceptions, these parallel upward trends suggest a structural evolution in fashion imagery toward slightly fuller physiques, while the overall ideal remains narrow relative to population norms.
\subsection{The influence of outliers}

\begin{figure}[h!]
    \centering
    \includegraphics[width=\textwidth]{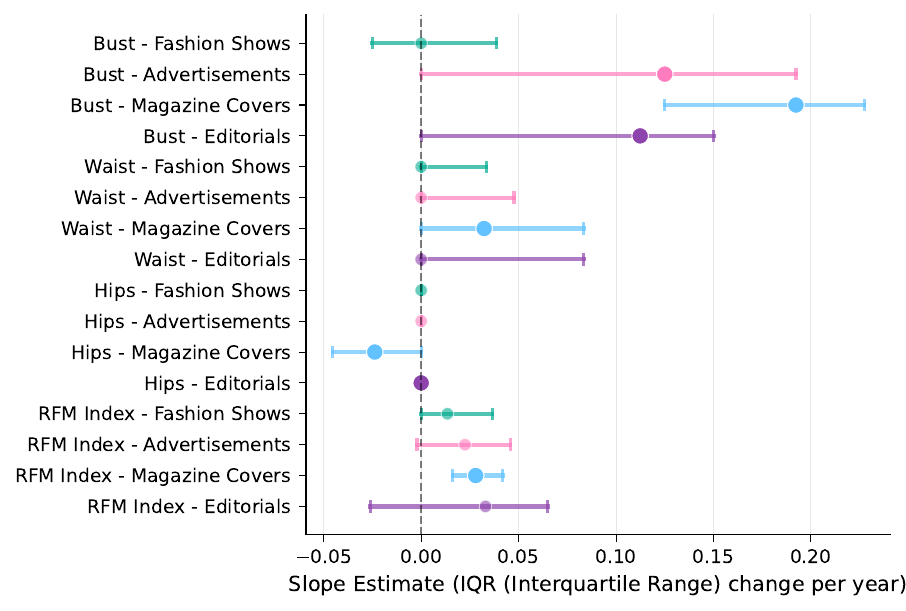}  
    \caption{Interquartile range trends and slope estimates.}
    \label{fig:si-forest-plot-iqr}
\end{figure}

\begin{figure}[h!]
    \centering
    \includegraphics[width=\textwidth]{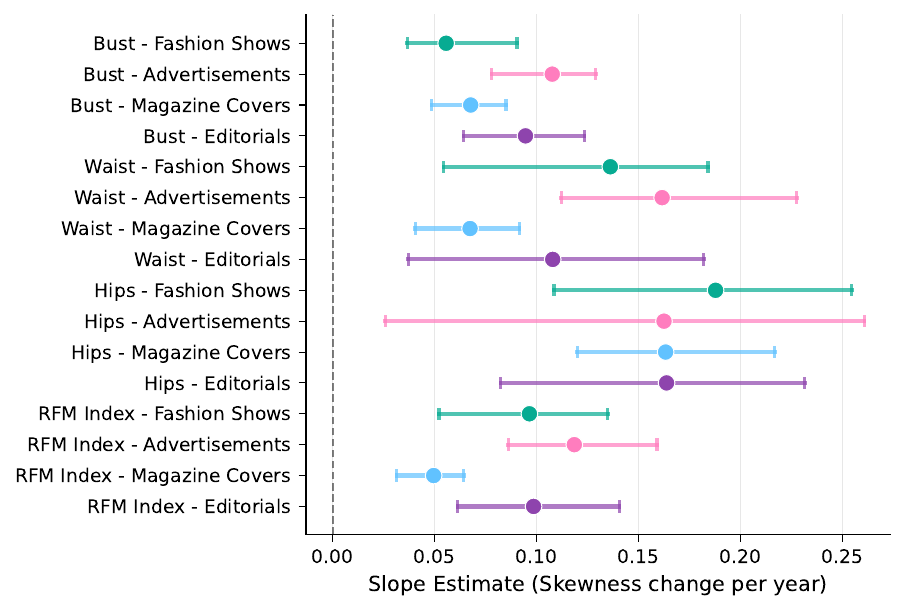}
    \caption{Skewness trends and slope estimates.}
    \label{fig:si-forest-plot-skew}
\end{figure}

\begin{figure}[h!]
    \centering
    \includegraphics[width=\textwidth]{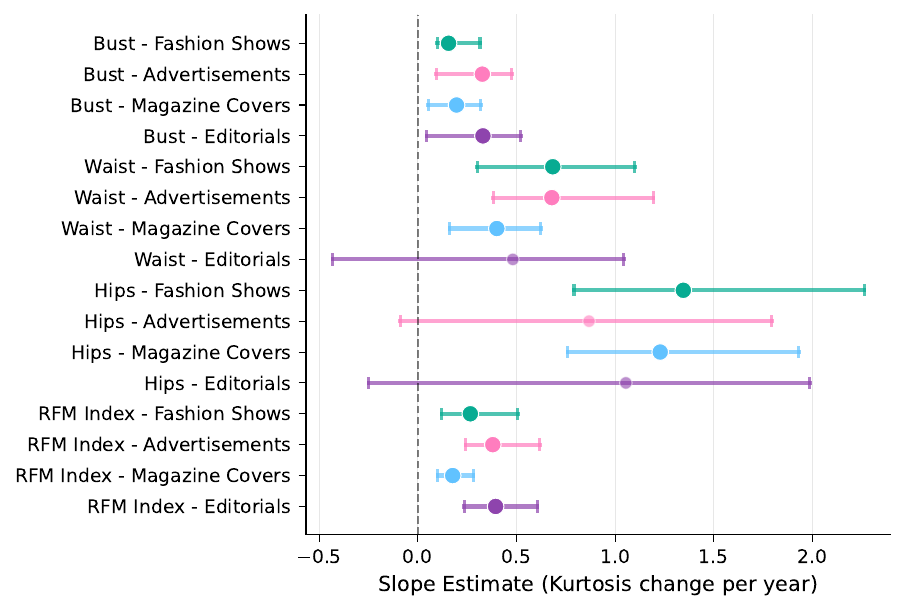}
    \caption{Kurtosis trends and slope estimates.}
    \label{fig:si-forest-plot-kurtosis}
\end{figure}

Distributional diagnostics (Figs. \ref{fig:si-forest-plot-iqr}--\ref{fig:si-forest-plot-kurtosis}) provide a detailed view of how the shapes of body measurement distributions evolve over time across media types.

Slope estimates for interquartile range (IQR) trends are generally close to zero across bust, waist, hips, and RFM (Fig.~\ref{fig:si-forest-plot-iqr}).
This indicates that the central 50\% of values remain relatively stable over time, regardless of source (fashion shows, advertisements, magazine covers, or editorials).
A few measurement-media combinations show small positive slopes (e.g., bust and waist in magazine covers), but confidence intervals are wide and mostly overlap zero, supporting the conclusion that the typical range of body sizes has not broadened.
By contrast, skewness shows consistent positive slopes for bust, waist, and hips (Fig.~\ref{fig:si-forest-plot-skew}), reflecting heavier right tails.
These effects are particularly pronounced in fashion shows and advertisements, suggesting that the upper extremes of body measurements have become more prominent over time.
RFM skewness trends are weaker and often close to zero, confirming that proportionality measures remain largely symmetric.
Kurtosis slopes are also predominantly positive for bust, waist, and hips (Fig.~\ref{fig:si-forest-plot-kurtosis}), indicating the emergence of heavier-tailed distributions.
The strongest increases are again seen in fashion shows and advertisements, while editorials show more muted or uncertain changes.
For the RFM index, most slopes remain near zero or only slightly positive, consistent with the notion that proportional measures are less affected by outliers.
Taken together, these diagnostics demonstrate that the apparent rise in diversity is not due to an overall widening of the central distribution but rather to changes in the tails.
Outliers, particularly larger body measurements, increasingly shape the distributions in fashion media, whereas indices of proportionality remain comparatively stable.
\newpage
\subsection{Multidimensional analysis: principal component analysis}
\label{si:pca}

\begin{figure}[htb]
    \centering
    \includegraphics[width=1\linewidth]{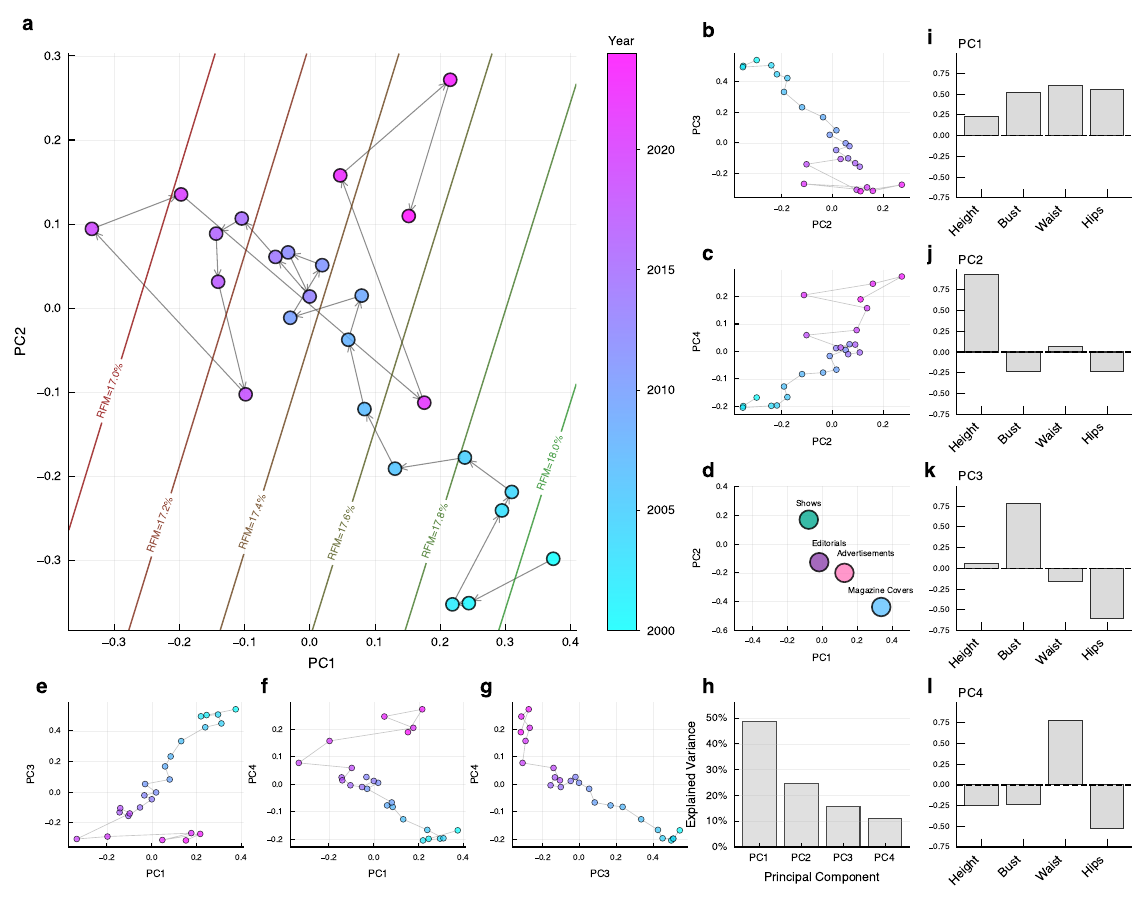}
      \caption[{Temporal evolution of body measurements in PCA space.}]{\textbf{Temporal evolution of body measurements in PCA space.}
      \textbf{(a)} Main trajectory: PC1-PC2 showing yearly means from 2000-2024 with directional arrows and RFM isolines (17.0-18.0\%).
      Colormap indicates temporal progression from early (blue) to recent years (pink).
      \textbf{(b-c)} Secondary projections: PC2-PC3 and PC2-PC4 reveal additional temporal patterns beyond the primary PC1-PC2 plane.
      \textbf{(d)} Career event averages: distinct positioning of \textcolor{teal}{fashion shows}, \textcolor{softpink}{advertisements}, \textcolor{lightblue}{magazine covers}, and \textcolor{purple}{editorials} in PC1-PC2 space demonstrates body measurement variation across fashion and media contexts.
      \textbf{(e-g)} Complementary PC combinations (PC1-PC3, PC1-PC4, PC3-PC4) showing consistent temporal trajectories across all principal component dimensions.
      \textbf{(h)} Explained variance: PC1-PC4 account for cumulative variance with PC1 and PC2 dominating.
      \textbf{(i-l)} Component loadings: waist and hips drive PC1 (body composition), height dominates PC2 (stature), bust and waist contribute to PC3-PC4 (secondary shape variation).
      (see Supplementary: Section \ref{si:pca}).}
\label{fig:si-pca}
\end{figure}

To account for multivariate body shape, we perform a principal components analysis (PCA) using height, bust, waist and hips.
Figure \ref{fig:si-pca} shows the trajectories of the yearly mean value in the PC1-PC2 space.
We observe that from 2000 to 2020 the mean of all measurements evolves roughly orthogonally to the RFM isolines, in the same way as the US models and US general population are separated by RFM isolines in Figure 2 of the manuscript.
These results show that the RFM is a particularly relevant quantity as it emerges as the main variable of evolution and separation.
We can also interpret the principal components. The first principal component (Fig.~\ref{fig:si-pca}i) measures the volume of models.
Indeed, all base variable loadings are positive but height has a smaller value as it is a one-dimensional variable compared to the three others, which are two-dimensional circumferences (bust, waist, hips).
The second principal component primarily captures height and explains the direction of the RFM isolines.
Figure \ref{fig:si-pca}e,f,g illustrate the increased representation of plus size models since 2020.

\clearpage
\newpage

\subsection{Network Analysis}

We model the brand--model relation as a weighted bipartite graph
\(G=(\mathcal{M},\mathcal{B},\mathcal{E})\), where \(\mathcal{M}\) and \(\mathcal{B}\) denote models and brands (fashion houses and magazines), respectively.
An edge \(e_{mb}\in\mathcal{E}\) carries weight \(w_{mb}\in\mathbb{R}_{+}\), the number of appearances of model \(m\) for brand \(b\) during the study period.
The weighted degree of node \(v\) is
\[
k_v=\sum_{u\in\mathcal{N}(v)} w_{vu},
\]
it corresponds to the number of works for a model or the number of events for a brand.
To quantify hierarchical influence, we compute eigenvector centrality on one-mode projections with fractional co-appearance weights (each event contributes unit mass across co-participants) \cite{Clauset2015SciAdv}.
For the model projection \(G_{\mathcal{M}}\),
\[
\tilde w_{mm'}=\sum_{b\in\mathcal{B}}\frac{w_{mb}\,w_{m'b}}{\sum_{m''\in\mathcal{M}} w_{m''b}}\!,
\]
and centralities \(c_v\) satisfy
\[
c_v=\lambda^{-1}\sum_{u\in\mathcal{N}_{\mathcal{M}}(v)} \tilde w_{vu}\,c_u,
\qquad \text{normalized so } \sum_{v\in\mathcal{M}} c_v=1.
\]
Brand prestige is computed analogously on the brand projection \(G_{\mathcal{B}}\) with
\[
\tilde w_{bb'}=\sum_{m\in\mathcal{M}}\frac{w_{mb}\,w_{mb'}}{\sum_{b''\in\mathcal{B}} w_{mb''}}\!.
\]

Tiering by appearance-weighted percentiles.
Because elite actors generate a disproportionate share of appearances, tier cutpoints are defined on appearance-weighted (edge-weighted) centrality distributions rather than on unweighted actor counts.
Let \(s_v:=k_v\) denote the appearance volume of actor \(v\). Define the weighted CDF
\[
F^{(s)}_c(t)=\frac{\sum_{v:\,c_v\le t} s_v}{\sum_{v} s_v},
\]
and the weighted quantile \(Q^{(s)}_{p}(c)=\inf\{t:\,F^{(s)}_c(t)\ge p\}\).
Actors are assigned to four ordered tiers by weighted percentile bands (share of total appearances):
\[
\mathcal{T}_{\mathrm{Elite}}=\{v:\,c_v\ge Q^{(s)}_{0.90}(c)\},\quad
\mathcal{T}_{\mathrm{High}}=\{v:\,Q^{(s)}_{0.50}(c)\le c_v<Q^{(s)}_{0.90}(c)\},
\]
\[
\mathcal{T}_{\mathrm{Mid}}=\{v:\,Q^{(s)}_{0.10}(c)\le c_v<Q^{(s)}_{0.50}(c)\},\quad
\mathcal{T}_{\mathrm{Low}}=\{v:\,c_v<Q^{(s)}_{0.10}(c)\},
\]
corresponding to 90-100 (Elite), 50-90 (High), 10-50 (Mid), and 0-10 (Low) percentiles by appearance mass.
Unless stated otherwise, downstream analyses condition on these tiers while retaining the continuous \(c_v\) as covariates in regression-based robustness checks.
\clearpage
\newpage

\subsection{Intersectionality}
\label{si:intersec-fits}

\begin{figure}[h!]
    \centering
    \includegraphics[width=0.7\textwidth]{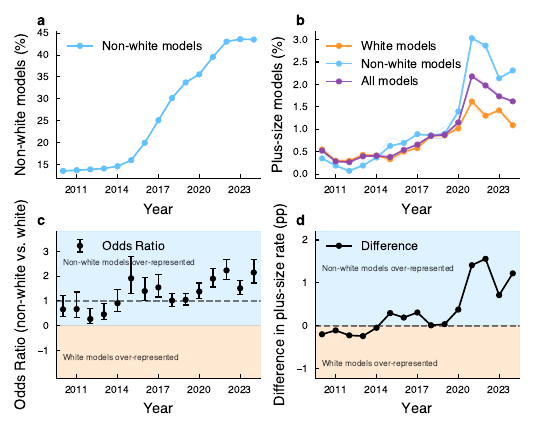}
    \caption[{Intersectionality results with plus-size models defined with RFM}]{\textbf{Intersectionality results with plus-size model defined with RFM}
    \textbf{(a)} Evolution of the proportion of non-White models.
    \emph{Methods: Ethnicity Attribution}.
    \textbf{(b)} Evolution of the proportion of plus-size model per ethnicity (White and non-White) (here plus-size is define as Top 1\% RFM values).
    \textbf{(c)} Odds ratio of being a plus-size model for the two populations.
    It represents how much more likely non-White models are to be plus-size compared to White models.
    \textbf{(d)} Difference between the proportion of plus-size model per ethnicity.
    non-White models are almost always over-represented among the plus-size models.} 
    \label{fig:si-intersectional-rfm}
\end{figure}

In this paragraph we show that the intersectionality we observe (non-White models being over-represented in the population of plus-size models) is robust to the definition of the plus-size group.
Instead of using the dress size (US dress size >12), we use the RFM (see Section Relative Fat Mass).
Here, We define plus-size models as those falling within the top 1\% of the RFM distribution. .
We find that the intersectional pattern persists under this alternative metric (Fig.~\ref{fig:si-intersectional-rfm}).
From 2015 to 2024, odds ratios consistently exceed 1, confirming that non-White models maintain a significantly higher likelihood of being classified as plus-size.

\begin{figure}[h!]
    \centering
    \includegraphics[width=\textwidth]{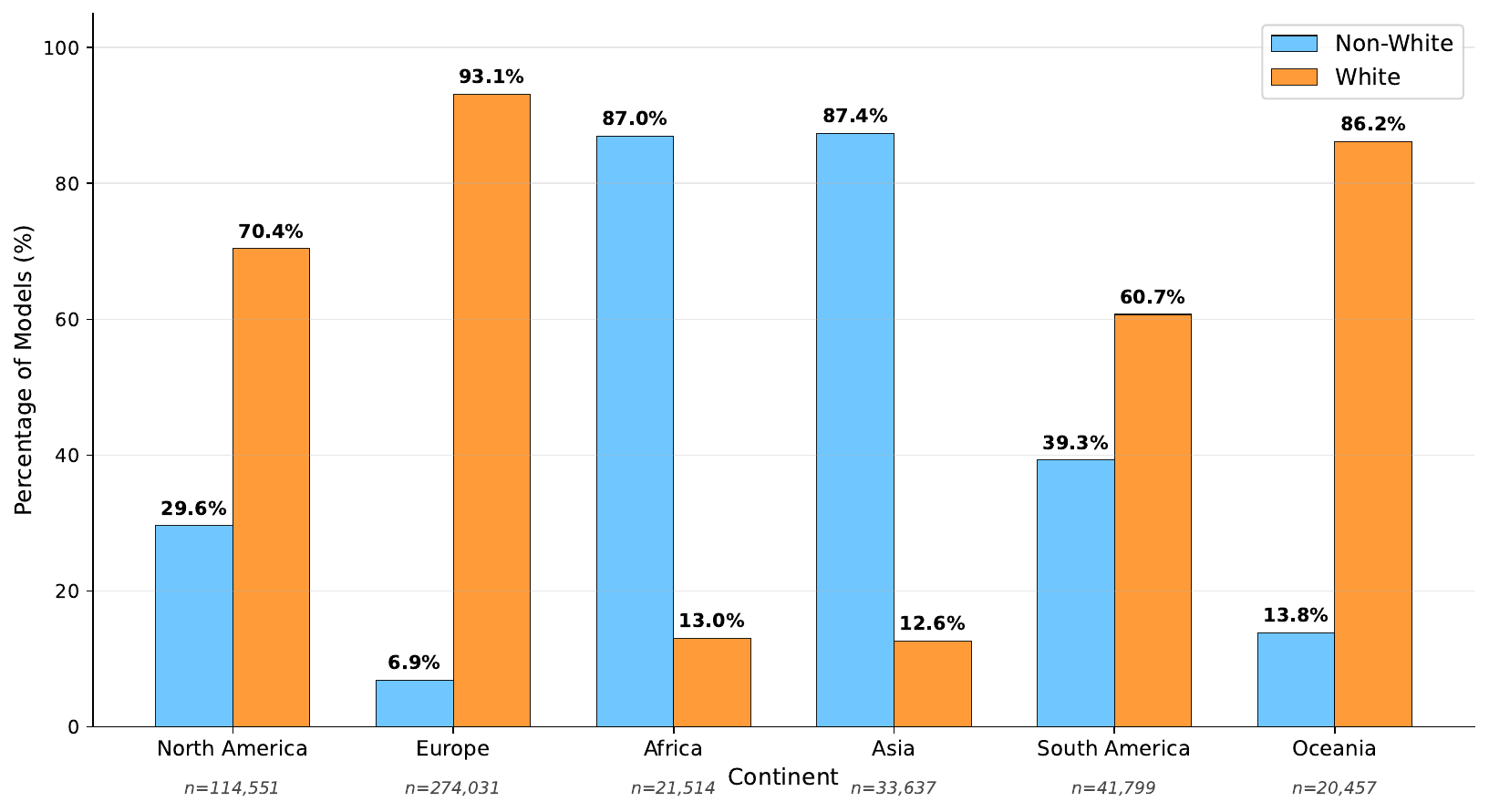}
    \caption[Ethnicity by continent]{Ethnicity distribution by continent} 
    \label{fig:si-intersectional-ethnicity}
\end{figure}

Figure \ref{fig:si-intersectional-ethnicity} shows the distribution of non-White and White models by continent.

\begin{figure}[h!]
    \centering
    \includegraphics[width=\textwidth]{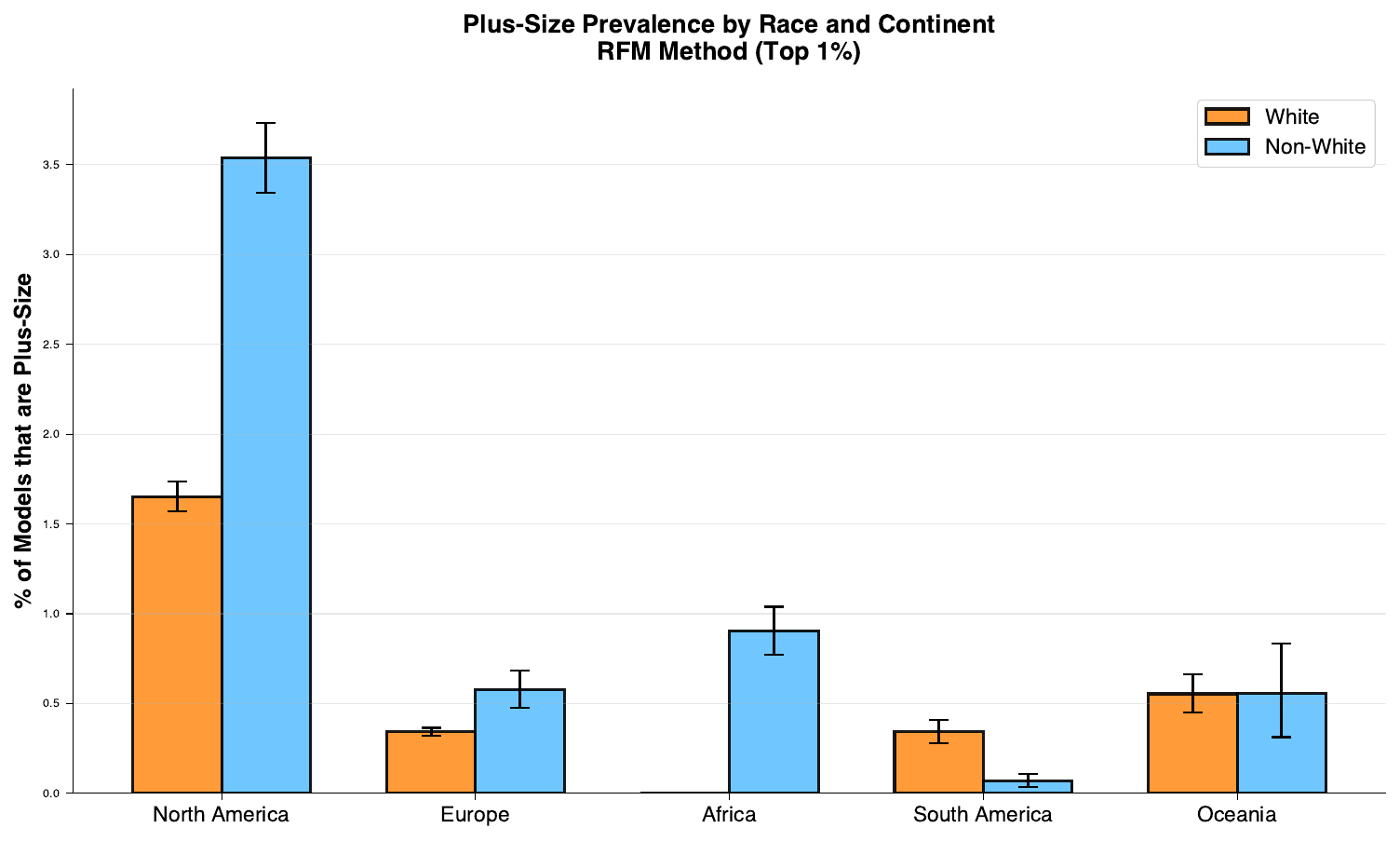}
    \caption[Intersectionality by continent (RFM)]{Proportion of plus-size model per ethnicity and continent} 
    \label{fig:si-intersectional-ethnicity-continent-rfm}
\end{figure}

\begin{figure}[h!]
    \centering
    \includegraphics[width=\textwidth]{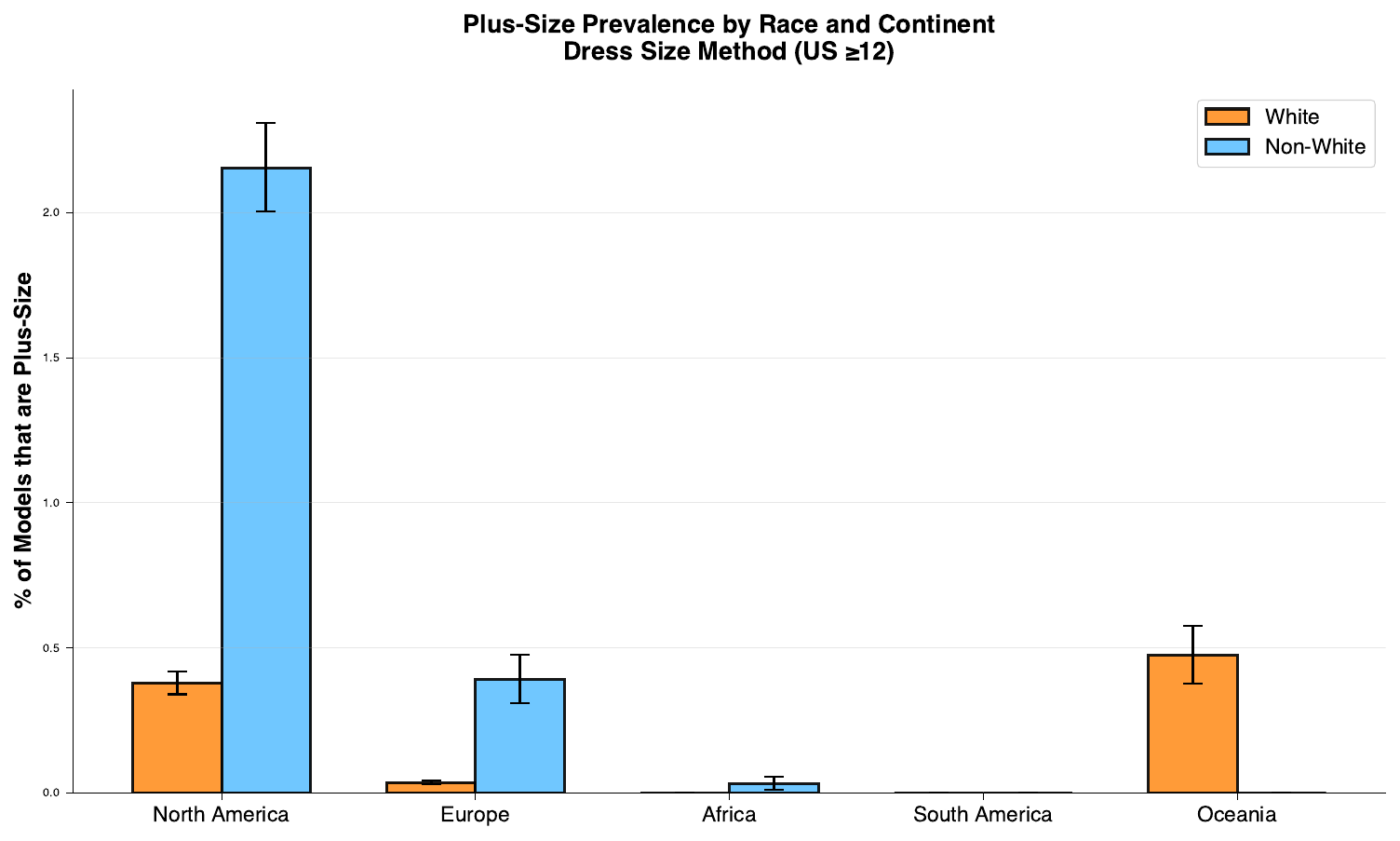}
    \caption[Ethnicity by continent (Dress size)]{Proportion of plus-size model per ethnicity and continent} 
    \label{fig:si-intersectional-ethnicity-continent-dress}
\end{figure}

To assess the geographic generalizability of the intersectional patterns documented in the main text, we decompose plus-size prevalence by continent and ethnicity (Figures~\ref{fig:si-intersectional-ethnicity-continent-dress} and \ref{fig:si-intersectional-ethnicity-continent-rfm}).
The over-representation of non-white models in plus-size categories holds consistently across continents with sufficient sample sizes.
In North America, non-white models exhibit plus-size rates of 2.15\% (dress size method) and 3.54\% (RFM method) compared to 0.38\% and 1.65\% for white models respectively, based on 126,721 model-work records.
Europe shows similar patterns (n=309,957), with non-white rates of 0.39\% and 0.58\% versus white rates of 0.04\% and 0.34\%.
The  consistency of these disparities across major fashion markets—despite different overall prevalence levels—demonstrates that the intersectionality of race and body size operates as a global industry pattern rather than a region-specific phenomenon.

\clearpage
\newpage

\subsection{Distributional Evidence of Policy Impact}
\label{si:sec-did-distribution}

\begin{figure}[h!]
    \centering
    \includegraphics[width=\textwidth]{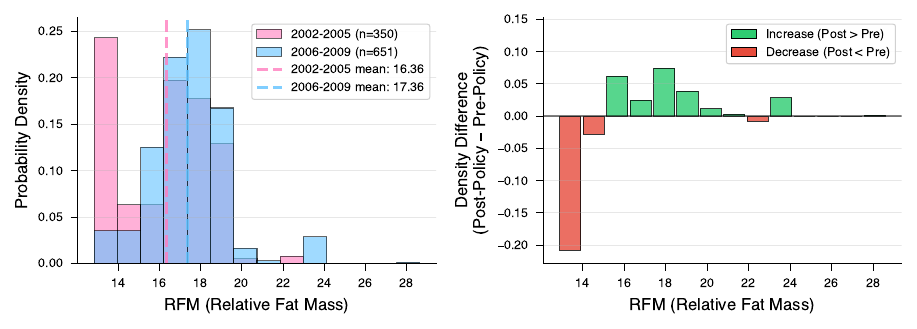}
    \caption[Distributional impact of the 2006 Milan BMI floor.]{\textbf{Distributional impact of the 2006 Milan BMI floor on model body composition.}
    \textbf{(Left)} RFM distributions at Milan Fashion Week before (2002--2005, $n=350$) and after (2006--2009, $n=651$) the policy intervention.
    Dashed lines indicate period means. The distribution shifts rightward, with mean RFM increasing from 16.36 to 17.36.
    \textbf{(Right)} Density difference (post-policy minus pre-policy). Red indicates decreased density; green indicates increased density.
    The policy reduced the prevalence of very low RFM values (left tail) while increasing density at moderate RFM levels, consistent with lower-tail pruning rather than a uniform distributional shift.}
    \label{fig:si-milan-rfm-distribution}
\end{figure}

In this section, we provide additional information regarding the policy interventions, presenting raw distributional shifts and time series data to characterize changes in model body composition.
In Milan, the introduction of the BMI threshold ($\ge 18.5$ kg m$^{-2}$) coincided with a large change of the body-size distribution.
Comparing the pre-policy (2002--2005) and post-policy (2006--2009) periods, we observe a large rightward shift in the Relative Fat Mass (RFM) distribution (Supplementary Figure~\ref{fig:si-milan-rfm-distribution}).
The density difference reveals a distinct pruning effect: the probability mass in the extreme lower tail (RFM $< 15$) is reduced, while density increases in the $16-20$ RFM range.
This indicates that the policy mechanism effectively targeted underweight outliers rather than merely shifting the population mean through a general trend.

\begin{figure}[h!]
    \centering
    \includegraphics[width=\textwidth]{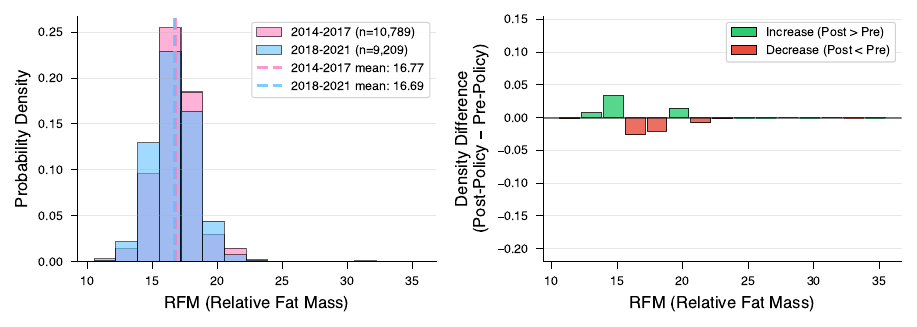}
    \caption[Distributional impact of the 2017 French medical-certificate law.]{\textbf{Distributional impact of the 2017 French medical-certificate law on model body composition.}
    \textbf{(Left)} RFM distributions at Paris Fashion Week before (2014--2017, $n=10{,}789$) and after (2018--2021, $n=9{,}209$) the policy intervention.
    Dashed lines indicate period means. The distributions are nearly identical, with mean RFM unchanged (16.77 vs.\ 16.69).
    \textbf{(Right)} Density difference (post-policy minus pre-policy). Unlike Milan (Fig.~\ref{fig:si-milan-rfm-distribution}), the density difference shows no systematic pattern: changes are small in magnitude and do not concentrate in the lower tail, indicating that the medical-certificate requirement without explicit numeric thresholds produced no detectable effect on body-size representation.}
    \label{fig:si-paris-rfm-distribution}
\end{figure}

Conversely, the 2017 French regulation, which required medical certificates without specifying numeric cutoffs, produced no detectable distributional shift.
The RFM distributions for Paris Fashion Week before (2014--2017) and after (2018--2021) the policy implementation are statistically indistinguishable (Supplementary Figure~\ref{fig:si-paris-rfm-distribution}).
The mean RFM remained effectively constant (16.77 vs. 16.69), and the density difference fluctuates around zero with no systematic reduction in the lower tail.
This suggests that non-numeric health requirements did not alter casting practices regarding underweight models.

\begin{figure}[h!]
    \centering
    \includegraphics[width=\textwidth]{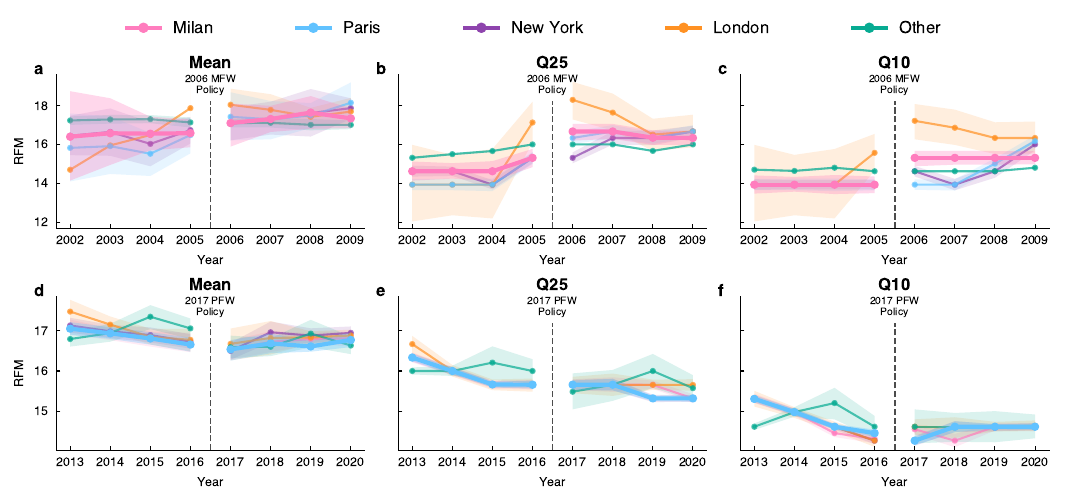}
    \caption[Temporal evolution of model RFM by city.]{\textbf{Temporal evolution of model RFM by city during regulatory interventions.}
    Evolution of the Mean, 25th percentile (Q25), and 10th percentile (Q10) of Relative Fat Mass (RFM) across major fashion capitals.
    \textbf{(a-c)} Trends around the 2006 Milan BMI floor. Milan (pink) exhibits a distinct upward discontinuity in 2006, diverging from the stable or declining trends of control cities (New York, Paris, London).
    The effect is most pronounced in the lower quantiles (Q10, Q25), confirming the policy's efficacy in addressing the lower tail of the distribution.
    \textbf{(d-f)} Trends around the 2017 Paris medical certificate law. Paris (blue) shows no deviation from the general industry trend (Other, green), with lower quantiles remaining stable or declining, indicating a null effect of the policy relative to the control group.}
    \label{fig:si-city_regulation}
\end{figure}

To contextualize these shifts, we analyze the temporal evolution of RFM statistics (Mean, Q25, Q10) across major fashion capitals (Supplementary Figure~\ref{fig:si-city_regulation}).
The time series confirm that the 2006 discontinuity is specific to Milan.
While New York, London, and Paris (Control) exhibit stable or declining RFM trends during the mid-2000s, Milan shows a sharp divergence in 2006, particularly in the 10th and 25th percentiles.
In contrast, during the 2017 intervention period, Paris tracks closely with the "Other" cities, exhibiting no differential break in trend.
This comparative view reinforces the finding that numeric thresholds (Milan) successfully decoupled local casting standards from the global industry trend towards thinness, whereas administrative requirements (Paris) did not.

\clearpage
\newpage

% PRINT SUPPLEMENTARY BIBLIOGRAPHY
\bibliographystyle{unsrt}
\bibliography{bibliography,bibliography_supplementary}

\end{document}